\documentclass[%twocolumn,
showpacs,preprintnumbers,amsmath,amssymb]{revtex4}\usepackage{amsmath}
\usepackage{graphicx}
\usepackage{subfig}
\usepackage{hyperref}
\usepackage{amssymb}
\usepackage[english]{babel}
\usepackage{epsfig}

\def\be{\begin{equation}}
\def\ee{\end{equation}}
\def\bea{\begin{eqnarray}}
\def\eea{\end{eqnarray}}

\def\nn{\nonumber \\}
\def\e{{\rm e}}

\begin{document}
%%%%%%%%%%%%%%%%%%%%%%%%%%%%%%%%%%%%%%%%%%%%
%%%%%%%%%%%%%%%%%%%%%%%%%%%%%%%%%%%%%%%%%%%%

\tolerance=5000

\title{Parametrizing the transition to the phantom epoch with Supernovae Ia and Standard Rulers}
\author{Iker Leanizbarrutia$^{a}$ and Diego S\'aez-G\'omez$^{a,b}$}
\affiliation{$^{a}$ Fisika Teorikoaren eta Zientziaren Historia Saila, Zientzia eta Teknologia Fakultatea,\\
Euskal Herriko Unibertsitatea, 644 Posta Kutxatila, 48080 Bilbao, Spain \\
$^{b}$Astrophysics, Cosmology and Gravity Centre (ACGC), and\\
Department of Mathematics and Applied Maths, University of Cape Town, Rondebosch 7701, Cape Town, South Africa}

\begin{abstract}
The reconstruction of a (non)canonical scalar field  Lagrangian from the dark energy Equation of State (EoS) parameter is studied, where it is shown that any EoS parametrization can be well reconstructed in terms of scalar fields.  Several examples of EoS parameters are studied and the particular scalar field Lagrangian is reconstructed. Then,  we propose some new parametrizations that may present a (fast) transition to a phantom dark energy EoS (where $w_{DE}<-1$) and  the scalar field Lagrangian is also reconstructed numerically. Furthermore, the properties of these parametrizations of the dark energy EoS are studied by using supernovae Ia data (HST Cluster Supernova Survey)  combined with Standard Ruler datasets [Cosmic Microwave Background (CMB) and Baryon Acoustic Oscillations (BAO)] and its comparison with the $\Lambda$CDM model is analyzed. Then, the best fit of the models is obtained, which provides some information about whether a phantom transition may be supported by the observations. In this regard, the crossing of the phantom barrier is allowed statistically but the occurrence of a future singularity seems unlikely. 
\end{abstract}

\maketitle

%%%%%%%%%%%%%%%%%%%%%%%%%%%%%%%%%%%%%%%%%%%%
\section{Introduction}
%%%%%%%%%%%%%%%%%%%%%%%%%%%%%%%%%%%%%%%%%%%%
In 1998 a deviation on the luminosity distance of Supernovae Ia (Sne Ia) was observed by two independent groups \cite{SN1}, a fact that was interpreted as the acceleration of the universe expansion. Later on,  other independent observations such as the Cosmic Microwave Background (CMB) \cite{WMAP}-\cite{Ade:2013zuv} or the Baryon Acoustic Oscillations (BAO) \cite{Eisenstein:2005su} have confirmed such hypothesis, which has been widely accepted by the scientific community since then. Then, a large number of candidates,  enclosed under the name of dark energy, have been proposed in order to understand the mechanism that produces such accelerating expansion (for a review on dark energy candidates see \cite{R-DE}).  The list of models includes a cosmological constant, canonical/phantom scalar fields \cite{Quintessence}, vector fields \cite{VT} or modifications of General Relativity (GR) \cite{modified-gravity}, among others. \\

Moreover, an interesting and useful approach for analyzing dark energy models is aimed to study the dark energy equation of state (EoS) as an effective description instead of reconstructing theoretical models. In this sense, dynamical EoS's that deviate from the cosmological constant have been widely studied, where perfect fluids with inhomogeneous EoS and redshift-dependent parameters have been proposed which may accomplish the late-time acceleration, and even the entire cosmological history by unifying the dark energy epoch and the inflationary phase (see Ref.~\cite{Nojiri:2005sr}).  Moreover, an effective description of the behavior of the dark energy EoS simplifies the fit of the free parameters while comparing with observational data, such that theoretical models, as modified gravities or  scalar-tensor theories, can be tested by using effective parametrizations of the EoS along the period of interest of the universe evolution. In this regard, several parametrizations of the dark energy EoS have been proposed over the last decade and its comparison with observational data has been studied (see Refs.~\cite{Huterer:2000mj}-\cite{Lazkoz:2007cc}). Some of these models lead to $\Lambda$CDM as the one with major statistical support but also other possibilities are allowed.  Furthermore,  the possibility that dark energy behaves as a phantom fluid, whose effective EoS parameter would turn out  $w_{DE}<-1$, has been also widely explored in the literature (see Ref.~\cite{phantom}) in spite of that such transition may lead to large instabilities in some particular phantom models \cite{vikman}. Such kind of EoS produces a phase of super-accelerating expansion that may end in a future singularity (for a classification of future singularities, see Refs.~\cite{Nojiri:2005sx}-\cite{barrow1}), whose analysis has attracted much interest, as may content important information on the structure of spacetime and its topology (see Ref.~\cite{barrow}). Hence, singular cosmologies have been  explored within several frameworks, including modified gravities (see Ref.~\cite{Nojiri:2008fk}). Furthermore, observations seem not to discard every phantom scenario and even some analysis highly support such possibility when studying carefully the observational data \cite{Lazkoz:2006gp}.\\

In the present paper, we present a reconstruction method for the action of a (non)canonical scalar field by just specifying the EoS parameter. Then, some examples of parametrizations of the dark energy EoS are reconstructed in terms of the scalar field.
Such reconstruction method may be extended to other  theoretical models as modified gravities to obtain the gravitational action from the dark energy EoS. The aim of such reconstruction is to provide a method to relate a phenomenological description, as the dark energy EoS, with the underlying theory that leads to such phenomenological behavior. Then, we  propose some new EoS  parametrizations  that experience fast changes, and which may give rise to fast crossings of the phantom barrier and eventually to the occurrence of a future singularity.  The best fit of the free parameters of the models are found by using some observational datasets  from Standard rulers  (CMB \cite{Spergel:2006hy} and BAO \cite{Eisenstein:2005su}) and Sne Ia \cite{Suzuki:2011hu}. The comparison of the results obtained by using each dataset is analyzed as well as  the comparison with the $\Lambda$CDM model.  The value of the relative matter density  $\Omega_m^0$ is found to be very close to the $\Lambda$CDM model, while the best fit of the EoS parameters does not discard the transition to the phantom epoch but disfavor the occurrence of future singularities.  \\

The paper is organized as follows: section \ref{reconstruction} deals with the reconstruction of scalar field models from the dark energy EoS. Then, section \ref{NewEoS} is devoted to the analysis of some new parametrizations of the EoS, where a preliminary study of the cosmological evolution and the occurrence of future singularities are analyzed.  Section \ref{ObsData} deals with the fit of the free parameters of the model with observational data and its comparison with $\Lambda$CDM. Finally, in section \ref{Conclusions},  we discuss the results of the paper.

%%%%%%%%%%%%%%%%%%%%%%%%%%%%%%%%%%%%%%%%%%
\section{Reconstructing scalar field models from the dark energy EoS}
\label{reconstruction}
%%%%%%%%%%%%%%%%%%%%%%%%%%%%%%%%%%%%%%%%%%

Let us consider a simple model with a scalar field besides  the matter content. Such an action can be expressed as follows \cite{Quintessence}
\begin{equation}
S=\int dx^{4}\sqrt{-g}\left[ \frac{1}{2\kappa^{2}}R 
 - \frac{1}{2} \gamma (\phi)
\partial_{\mu} \phi \partial^{\mu }\phi -V(\phi )+\mathcal{L}_{m}\right]\ ,
\label{ST1}
\end{equation}
where $ \kappa^2 =8\pi G$, $\mathcal{L}_m$ is the matter Lagrangian density whereas $\gamma(\phi)$ and $V(\phi )$ represent the kinetic term and the potential of the scalar field $\phi$ respectively.  By assuming a spatially
flat Friedmann-Lema\^itre-Robertson-Walker (FLRW) metric $ds^{2}=-dt^{2}+a^2(t) \sum_{i=1}^{3} dx_{i}^{2}$, the resulting  equations are given by
\be
H^{2} =\left(\frac{\dot{a}}{a}\right)^2= \frac{\kappa ^{2}}{3}\left( \rho _{m}
+\rho_{\phi}\right)\ , \quad \quad \dot H = -\frac{\kappa ^{2}}{2}\left(
\rho _{m}+p_{m}+\rho _{\phi }+p_{\phi }\right)\ ,
\label{ST2}
\ee
where 
\be
\rho _{\phi } = \frac{1}{2} \gamma (\phi )\, {\dot \phi}^{2}
+V(\phi)\ ,\quad \quad
p_{\phi } = \frac{1}{2} \gamma (\phi ) \, {\dot \phi}^{2}-V(\phi)\ ,
\label{ST3}
\ee
while the scalar field equation yields,
\be
\ddot{\phi}+3H\dot{\phi}+\frac{1}{2\gamma(\phi)}\left[\gamma'(\phi)\dot{\phi}^2+2V'(\phi)\right]=0\ .
\label{ST3a}
\ee
The matter content is described by a perfect fluid with a constant EoS $p_m=w_m\rho_m$,  such that the continuity equation $\dot{\rho}_m+3H(1+w_m)\rho_m=0$ can be easily solved leading to
\be
\rho_m=\rho_0a^{-3 (1+w_m)}=\rho_0 (1+z)^{3 (1+w_m)}\ ,
\label{ST4}
\ee
where $1+z=1/a$ is the redshift and $a_0=1$ is the value of the scale factor evaluated today. Furthermore,  the EoS parameter for the scalar field $\phi$ is defined as follows
\be
w_{\phi}=\frac{p_{\phi}}{\rho_{\phi}}=-1+\frac{\gamma(\phi)\dot{\phi}^2}{\frac{1}{2}\gamma(\phi)\dot{\phi}^2+V(\phi)}\ ,
\label{ST5}
\ee 
where $w_{\phi}$  depends in general on the cosmic time, or equivalently on the redshift. Note that any phantom transition gives rise to change the sign of the kinetic term $\gamma(\phi)$  as $\rho_{\phi}$ is defined positive. In addition, the assumption of the kinetic factor $\gamma(\phi)$ allows to redefine the scalar field, so that the scalar Lagrangian density can be easily reconstructed, as shown below.  Then,  introducing (\ref{ST4}-\ref{ST5}) in the FLRW equations (\ref{ST2}) and rewriting the equations in terms of the redshift instead of the cosmic time, the equation for the Hubble parameter reduces to
\begin{equation}
2H(1+z)H'-3 H^2 (1+w_{\phi}(z)) + 3 H^2_0 \Omega_m^0 (1+z)^{3(1+w_m)} w_{\phi} (z) = 0 \ ,
\label{ST6}
\end{equation}
where $\Omega_m^0 = \frac{\rho_{m0}}{3H_0^2/\kappa^2}$ is the relative matter density  and $H_0$ is the experimental value of the Hubble parameter evaluated today ($z=0$). In order to simplify the equation  (\ref{ST6}) the Hubble parameter can be 
redefined as $H(z)=H_0\ E(z)$. Then, the equation (\ref{ST6}) becomes
\begin{equation}
2E(z)(1+z)E'(z)- 3 E^2(z) (1+w_{\phi}(z)) + 3 \Omega_m^0 (1+z)^{3(1+w_m)} w_{\phi}(z) = 0\ .
\label{ST7}
\end{equation}
Hence, by specifying the EoS parameter (\ref{ST5}), the Hubble evolution $E(z)$ is obtained by solving the equation (\ref{ST7}). Note that in the case  $w_{\phi}=-1$, the kinetic term becomes null $\gamma(\phi)=0$ and the scalar field turns out constant,  such that the action is reduced to the $\Lambda$CDM model with a cosmological constant given by $2\Lambda=V_0$. Let us now consider the dynamical case, where the EoS parameter (\ref{ST5}) evolves with time. In such a case the following scalar potential and kinetic term are assumed,
%
%For simplicity we may rewrite the scalar field EoS (\ref{ST5}) in terms  of the redshift instead of the scalar field itself and consider an EoS parameter as follows
%\be
%w_{\phi}=-1+\tilde{w}_{\phi}(z)\ ,
%\label{ST8}
%\ee
%where $\tilde{w}_{\phi}$ would represent deviations from the $\Lambda$CDM model and is not constant in general. 
%Then, by assuming the following potential and  kinetic term 
\bea
\gamma(\phi)&=&-\frac{2}{\kappa^2}\frac{g'(\phi)}{\phi g(\phi)}-\frac{1+w_m}{g^2}\rho_{m0}\phi^{-(5+3w_m)}\ , \nn
V(\phi)&=&-\left(\frac{1-w_{\phi}(\phi)}{2(w_{\phi}(\phi)+1)}\right)\left(\frac{2}{\kappa^2}\phi g'(\phi)g(\phi)+\rho_{m0}(1+w_m)\phi^{-3(1+w_m)}\right)\ ,
\label{ST9}
 \eea
 which lead to the general solution,
 %the equation (\ref{ST6}) is solved,
\be
H=g\left(\frac{1}{1+z}\right)\ \quad \text{and} \quad \phi=\frac{1}{1+z}\ .
\label{ST10}
\ee
%Thus, by specifying the EoS parameter (\ref{ST5}), the corresponding scalar potential and kinetic term can be reconstructed, so that  scalar field Lagrangian (\ref{ST1}) can be effectively described by a particular EoS parametrization. \\
 Let us illustrate the above reconstruction by considering some examples. Firstly, the well known parametrization of the dark 
energy EoS suggested in \cite{Huterer:2000mj}, which is given by
\be
w(z) = w_0 + w_1 z, \quad w_1= \left(\frac{dw}{dz} \right)_{z=0} \ .
\label{ST11}
\ee
This parametrization, also called Linear Redshift Parametrization, initially proposed by Huterer and Turner 
in 2001 and by Weller and Albrecht in 2002, is only 
compatible with low redshift data ($z<1$) since grows linearly in redshift. In this case, the equation (\ref{ST7}) can be solved exactly by considering a pressureless matter fluid $w_m=0$, and it yields
\be
 H(z)=H_0\left(1+z\right)^{\frac{3}{2}(1-w_1)}\sqrt{\Omega_m^0(1+z)^{3w_1}+C_1\e^{3w_1(1+z)}(1+z)^{3w_0}}\ ,
 \label{ST12}
 \ee
 where $C_1$ is an integration constant. Then,  the kinetic term and the scalar potential (\ref{ST9}) can be obtained, where $g(\phi)=H(\frac{1-\phi}{\phi})$ given by (\ref{ST12}),
 \bea
 \gamma(\phi)=\frac{3C_1}{\kappa^2}\frac{w_1\left(1-\phi\right)+(1+w_0)\phi}{C_1\e^{\frac{3w_1}{\phi}}\phi^{3}+\Omega_m\phi^{3(1+w_0-w_1)}}\e^{\frac{3w_1}{\phi}}\ , \nn
 V(\phi)=\frac{3C_1H_0^2}{2\kappa^2}\left[w_1(-1+\phi)+(1-w_0)\phi\right]\e^{\frac{3w_1}{\phi}}\phi^{-4+3(w_1-w_0)}\ .
 \label{ST13}
 \eea
Hence the scalar field Lagrangian (\ref{ST1}) is fully reconstructed. Moreover, the best fit of the EoS parameters (\ref{ST11}) when set
with SNe Ia data are given by $w_0 = -1.4$ and $w_1 = 1.67$, \cite{Dicus:2004cp}, which presents a phantom transition which is well described by the kinetic term and scalar potential (\ref{ST13}), where $\gamma(\phi)$ changes its sign along the universe evolution. Let us consider now a slightly modified parametrization,
\be
w(z) = w_0 + w_1 z+w_2 z^2 \ ,
\label{ST14}
\ee
where a second order correction is included. Then, by solving the FLRW equations  (\ref{ST7}), the Hubble parameter is given by
\be
 H(z)=H_0\left(1+z\right)^{\frac{3}{2}(1-w_1)}\sqrt{\Omega_m^0(1+z)^{3w_1}+C_1\e^{\frac{3}{2}(1+z)[2w_1+w_2(z-3)]}(1+z)^{3(w_0+w_2)}}\ ,
\label{ST15}
\ee
And then, the kinetic term and the scalar potential yield
 \bea
 \gamma(\phi)=\frac{3C_1}{\kappa^2}\frac{w_2\left(1-\phi\right)^2+\phi\left[w_1+\phi(1+w_0-w_1)\right]}{C_1\e^{\frac{3(w_2+2w_1\phi)}{2\phi^2}}\phi^{4}+\phi^{3(1+w_0-w_1)}}\e^{\frac{3(w_2+2w_1\phi)}{2\phi^2}}\ , \nn
V(\phi)=-\frac{3C_1H_0^2}{2\kappa^2}\left[w_2(-1+\phi)^2+w_1\phi+(-1+w_0-w_1)\phi^2\right] \e^{\frac{3(w_2+(2w_1-4w_2)\phi)}{2\phi^2}}\phi^{-4+3(w_0-w_1+w_2)}\ .
 \label{ST16}
 \eea
Then, finally let us consider the following, also well known, parametrization \cite{Chevallier:2000qy},
\be
w(z)=w_0+w_1\frac{z}{1+z}\ .
\label{ST17}
\ee
Here the dark energy EoS tends to a constant for large redshifts while its dynamical behavior becomes important at small redshifts. Then, the Hubble parameter yields,
\be
H(z)=H_0^2 (1+z)^{3/2}\sqrt{\Omega_m+C_1(1+z)^{3(w_0+w_1)}\e^{\frac{3w_1}{(1+z)}}}\ .
\label{ST18}
\ee
And as in the previous examples, the kinetic term and the scalar potential (\ref{ST9}) which described the EoS parameter (\ref{ST17}) are reconstructed leading to
\bea
\gamma(\phi)=\frac{3C_1}{\kappa^2}\frac{1+w_0+w_1(-1+\phi)}{C_1\e^{3w_1\phi}+\Omega_m\phi^{3(w_0+w_1)}}\e^{3w_1\phi}\phi^{-2}\ , \nn
V(\phi)=\frac{3C_1H_0^2}{2\kappa^2}\left[1-w_0+w_1(-1+\phi)\right]\e^{3w_1\phi}\phi^{-3(1+w_0+w_1)}\ .
\eea
Hence, the reconstruction method explained above provides a way to get the underlying scalar field action for a particular dark energy EoS. However, note that in general, more complex EoS parametrizations would not lead to exact expressions for scalar field Lagrangian, but numerical resources are required. In the next section, a new parametrization that also may transit to the phantom epoch is proposed and the best fit is found by using Sne Ia, CMB and BAO data. The reconstruction of the scalar field Lagrangian is analyzed by using numerical methods.
 
 %%%%%%%%%%%%%%%%%%%%%%%%%%%%%%%%%%%%%%%%%%%%%%%%%%
 \section{Parametrizing the transition to the phantom epoch}
 \label{NewEoS}
 %%%%%%%%%%%%%%%%%%%%%%%%%%%%%%%%%%%%%%%%%%%%%%%%%%

Let us now assume a new parametrization for $w_{DE}(z)$ that may cross the phantom barrier 
 ($w < -1$) along the universe evolution,
\begin{equation}
w_{1}(z)=- 1+ w_{0}\left[ \tanh\left(z-z_0\right)-1\right]\ . 
\label{2.1}
\end{equation}
where $w_{0}$ and $z_0$ are free parameters. Specifically, $z_0$  
displaces the turning point of the function along the $z$ axis and $w_0$ 
controls the value of the EoS parameter when $z\leq z_0$, and indeed how far 
the phantom barrier is crossed and the time for the occurrence of future 
singularities, as discussed below. Note also that for $w_{0}<0$, there will 
not be phantom epoch as  $w_1>-1$ at any redshift.  \\

Moreover, the above parametrization can be slightly modified to become a 
transition parametrization centered around $w=-1$, leading to the second 
parametrization that is analyzed here,
\begin{equation}
w_2(z)=-1+ w_0 \tanh\left(z-z_0\right)\ ,  
\label{2.2}
\end{equation}
where in this case $w_0$ controls the width of the strip around $w=-1$ in which 
the parametrization can evolve. Nevertheless, both parametrizations behaves as 
$\Lambda$CDM  for $w_0=0$. 
Moreover, the EoS parameter (\ref{2.1})
tends to $w_1\sim-1$ at large redshifts and the model (\ref{2.2}) leads to $w_2\sim-1+w_0$, whereas  $w_1=-1-w_0$  and $w_2=-1$ at  $z=z_0$, the EoS transition point. Both parametrizations describe  deviations from $\Lambda$CDM. Furthermore, the reconstruction of the scalar field Lagrangian studied in the previous section can be applied to both EoS but numerical resources have to be applied in order to obtain the kinetic term and the scalar potential, since the FLRW equation (\ref{ST6}) does not lead to an exact solution in this case (see the Appendix for an exact solution using approximation methods). Then, for an illustrative purpose, the kinetic term and the scalar potential for a sample of the EoS models (\ref{2.1}) and (\ref{2.2}) are shown in Fig.~1. In both models, the scalar field rolls down along the potential at small redshifts when the dynamics of the scalar field becomes important, reaching a plato at the current time ($\phi=1$). The kinetic term presents also a similar asymptotic behavior. Consequently,  both models tend to a constant EoS asymptotically.\\

%Nevertheless, here we are interested to check whether some parametrization as (\ref{2.1}-\ref{2.2}), which can predict a phantom epoch along the universe evolution,  fit well with the observational data.  \\

\begin{figure}[h!]
  \centering
  \subfloat{\includegraphics[width=0.475\textwidth]{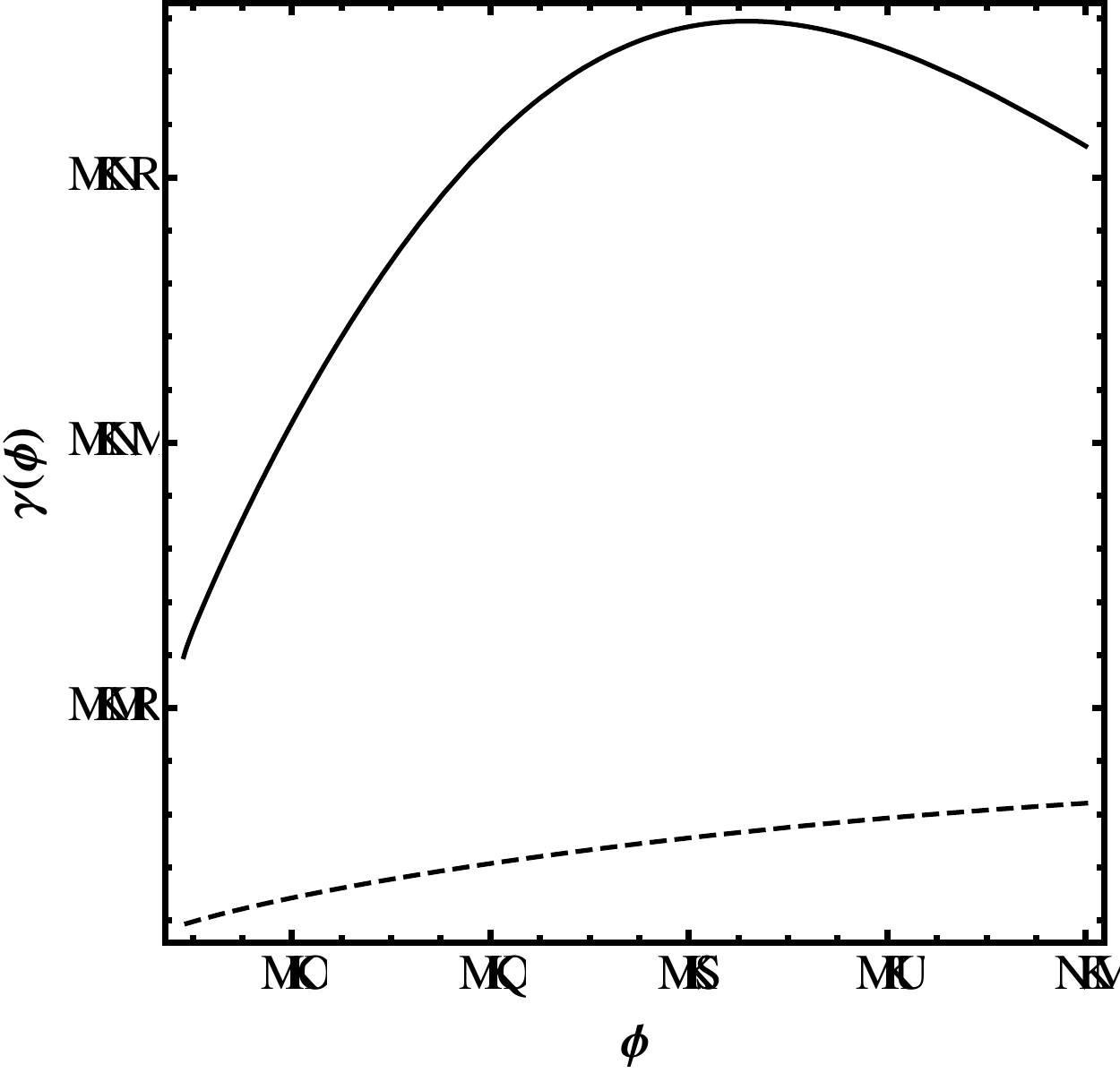}}
  \subfloat{\includegraphics[width=0.475\textwidth]{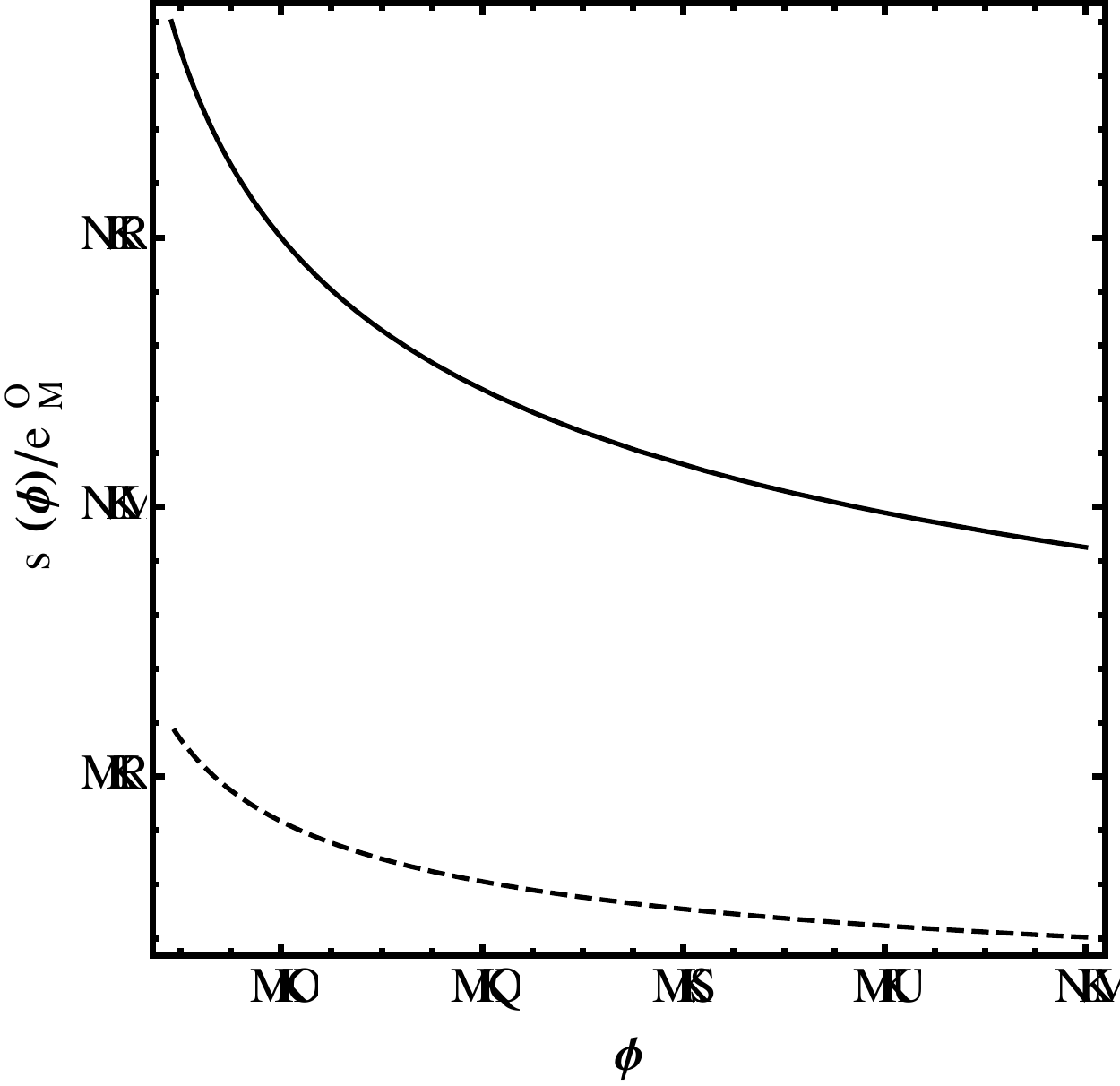}}
\caption{Reconstruction of the kinetic term and the scalar potential for the parametrization $w_1$ (solid line) and $w_2$ (dashed line). Here we have assumed $w_0=-0.05$ and $z_0=12.14$ for the model (\ref{2.1}) and $w_0=0.1$ and $z_0=-10$ for the model (\ref{2.2}).} \label{fig:1}
\end{figure}

Let us now analyze the cosmological evolution in a qualitative way for both models (\ref{2.1}) and (\ref{2.2}). By assuming  $\Omega_m^0=0.3$ and setting the initial conditions in order to fit $\Lambda$CDM  at $z=2000$, the equation (\ref{ST7}) is solved numerically for different values of the parameters $w_0$ and $z_0$, where we have assumed those values which approach closely to the $\Lambda$CDM model. Then, in Fig.~\ref{fig:1}, the evolution of the Hubble parameter is depicted, 
where the blue line corresponds to the $\Lambda$CDM model, whereas 
Fig.~\ref{fig:2} shows the evolution of the deceleration parameter. At large 
redshifts, the functions match the $\Lambda$CDM model as expected, whereas at 
small redshifts, where the dynamical behavior of the EoS parameters (\ref{2.1}) and (\ref{2.2})  becomes important, both the Hubble parameter as the deceleration may provide differences with respect the $\Lambda$CDM model, as shown in Figs.~\ref{fig:1}-\ref{fig:2}. 
%In addition, some combinations where both parameters have large values (close to considered maximum values), the resulting  $H(z)/H_0$ function presents a large peak at $z=0$, which directly excludes those values for possible fits with experimental data. Discarding those functions with anomalous behaviors, the Hubble and the deceleratio n parameters are depicted altogether with the $\Lambda$CDM model in Fig \ref{fig:1} and Fig. \ref{fig:2}.\\
%\begin{figure}[t]
%  \subfloat[First sub-figure\label{fig:1.figA}]{\includegraphics[width=0.35\textwidth]{images/h(z)1}}
%  \subfloat[First sub-figure\label{fig:1.figB}]{\includegraphics[width=0.35\textwidth]{images/h(z)2}}
%\caption{Evolution of the Hubble parameter for the models (\ref{2.1}) and (\ref{2.2}), where $w_0\in[-0.25,0.45]$ and $z_0\in[-100,400]$, and $\Lambda$CDM model (blue).} \label{fig:1}
%\end{figure}

\begin{figure}[h!]
  \centering
  \subfloat{\includegraphics[width=0.475\textwidth]{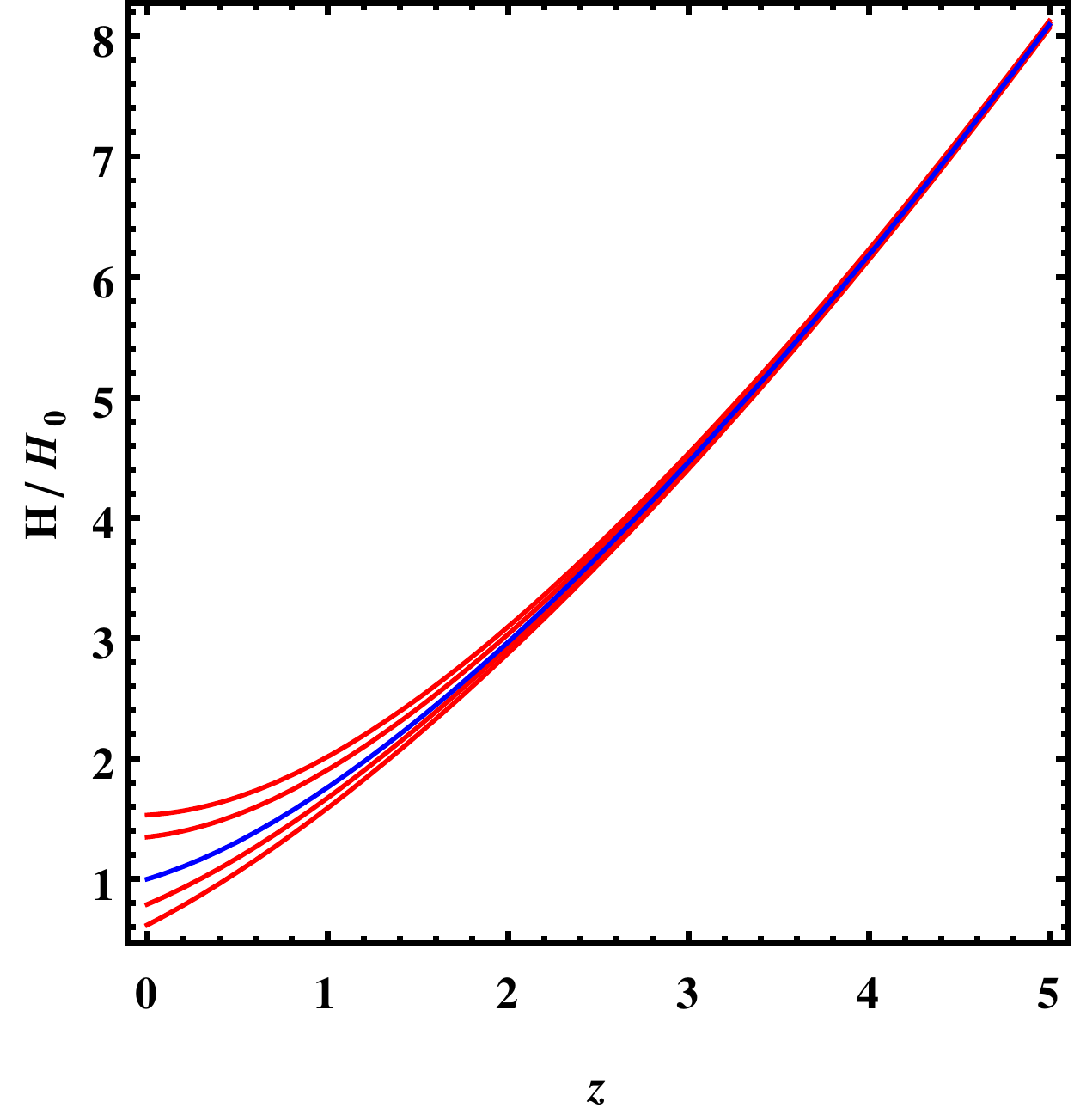}}
  \subfloat{\includegraphics[width=0.475\textwidth]{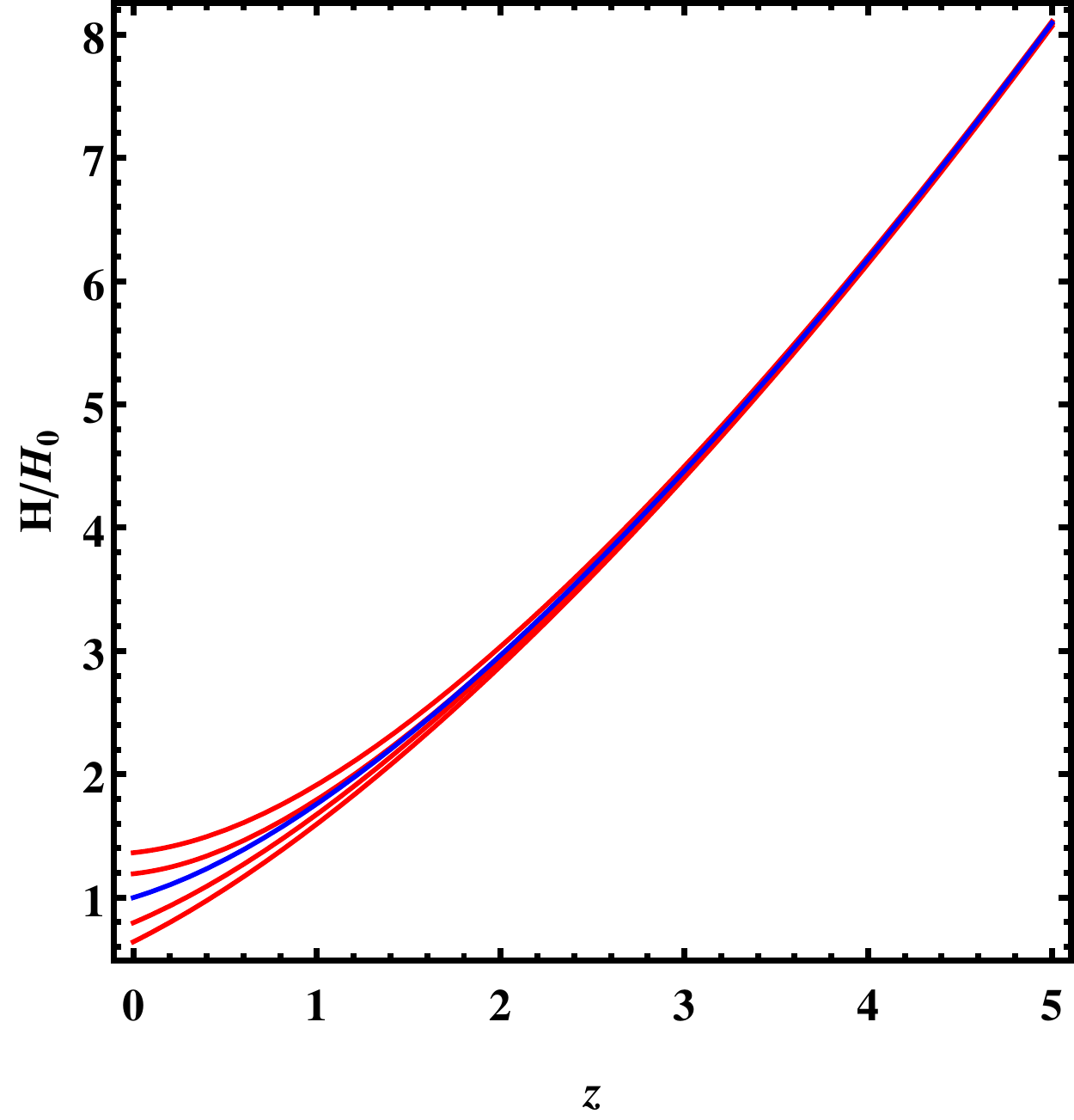}}
\caption{Evolution of the Hubble parameter assuming the EoS parametrizations (\ref{2.1}) and (\ref{2.2}) in comparison with $\Lambda$CDM  (blue line). For the first model (left panel), the parameter values are $(w_0,z_0)=\lbrace(0.05,34.28),(0.05,12.14),(-0.05,12.14),(-0.1,34.28)\rbrace$ from the upper to bottom. For the second model (right panel) the values in the same order are $(w_0,z_0)=\lbrace(0.1,12.14),(0.15,12.14),(0.1,-10),(0.15,-10)\rbrace$. } \label{fig:1}
\end{figure}

%We have also numerically computed the deceleration parameter, using the formula 
%(\ref{1.7}). The deceleration parameter is also depicted in Fig. \ref{fig:2} for 
%the same set of the free parameters $w_0$ and $z_0$.
\begin{figure}[h!]
  \centering
  \subfloat{\includegraphics[width=0.475\textwidth]{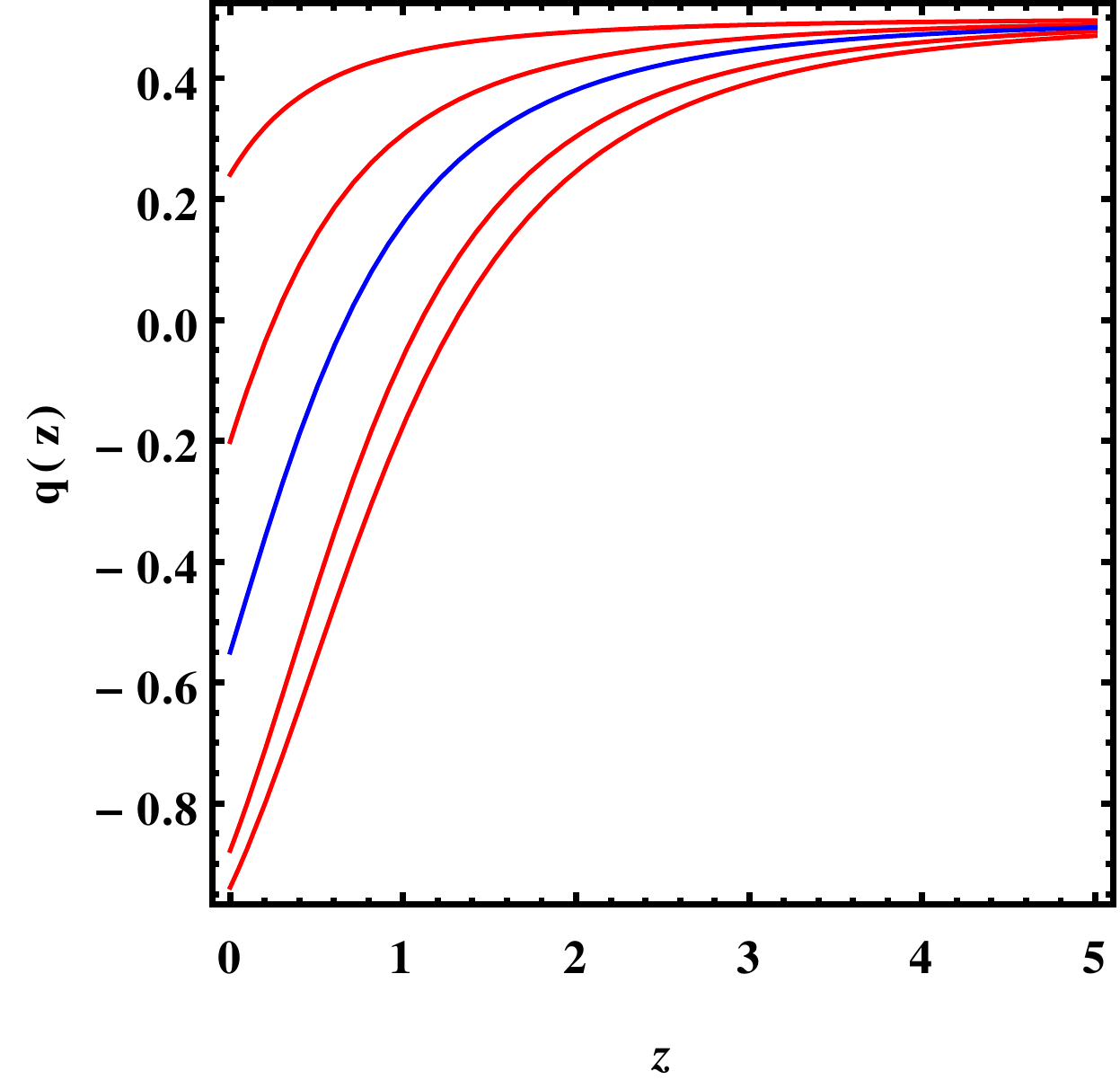}}
  \subfloat{\includegraphics[width=0.475\textwidth]{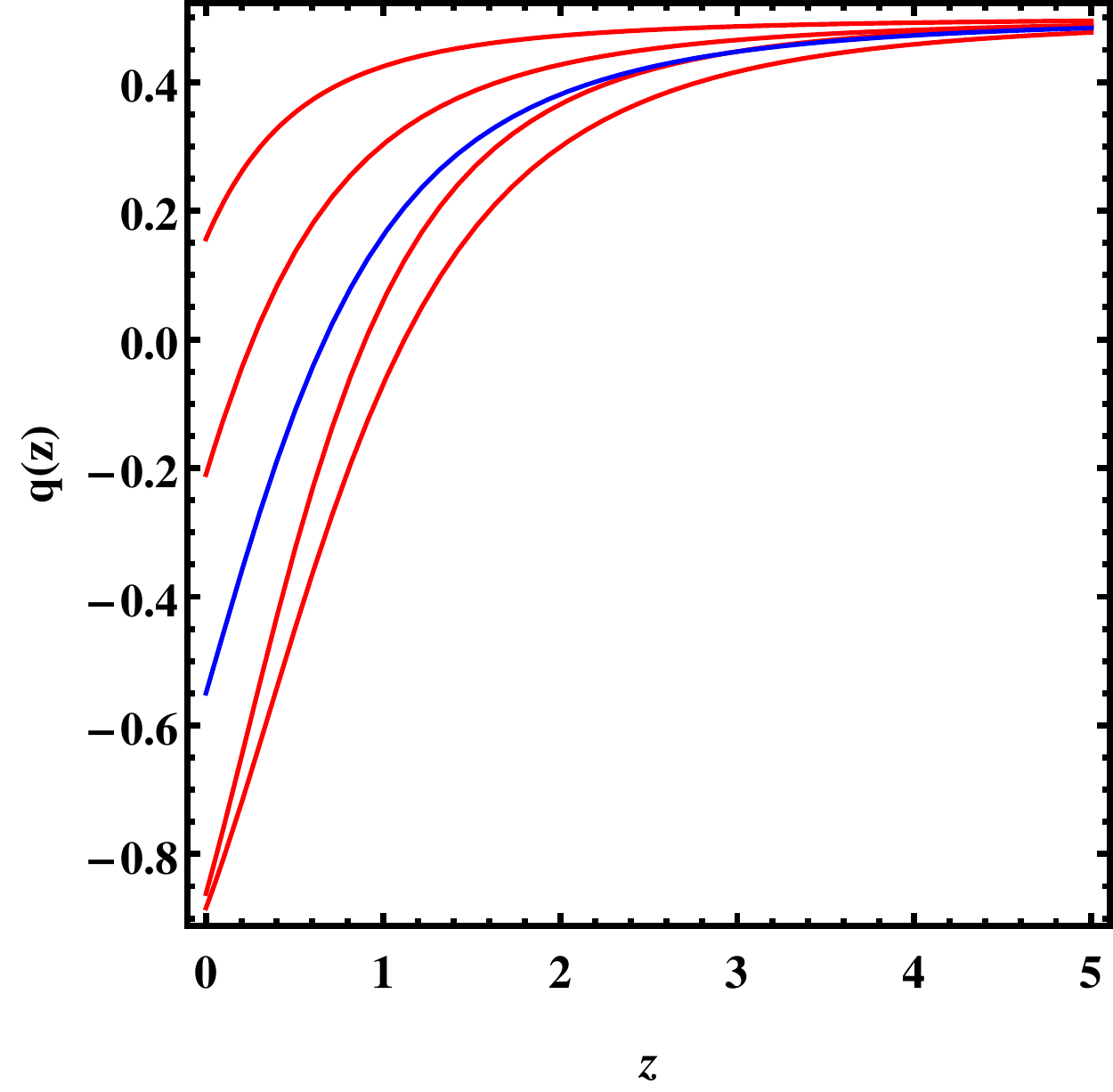}}
\caption{Evolution of the Hubble parameter for the EoS (\ref{2.1}) and (\ref{2.2}) in comparison with $\Lambda$CDM  (blue line). For the first model (left panel), the parameter values are $(w_0,z_0)=\lbrace(0.05,34.28),(0.05,12.14),(-0.05,12.14),(-0.1,34.28)\rbrace$ from the bottom to upper. For the second model (right panel) the values in the same order are $(w_0,z_0)=\lbrace(0.1,12.14),(0.15,12.14),(0.1,-10),(0.15,-10)\rbrace$.} \label{fig:2}
\end{figure}

\begin{figure}[h!]
  \centering
  \subfloat{\includegraphics[width=0.475\textwidth]{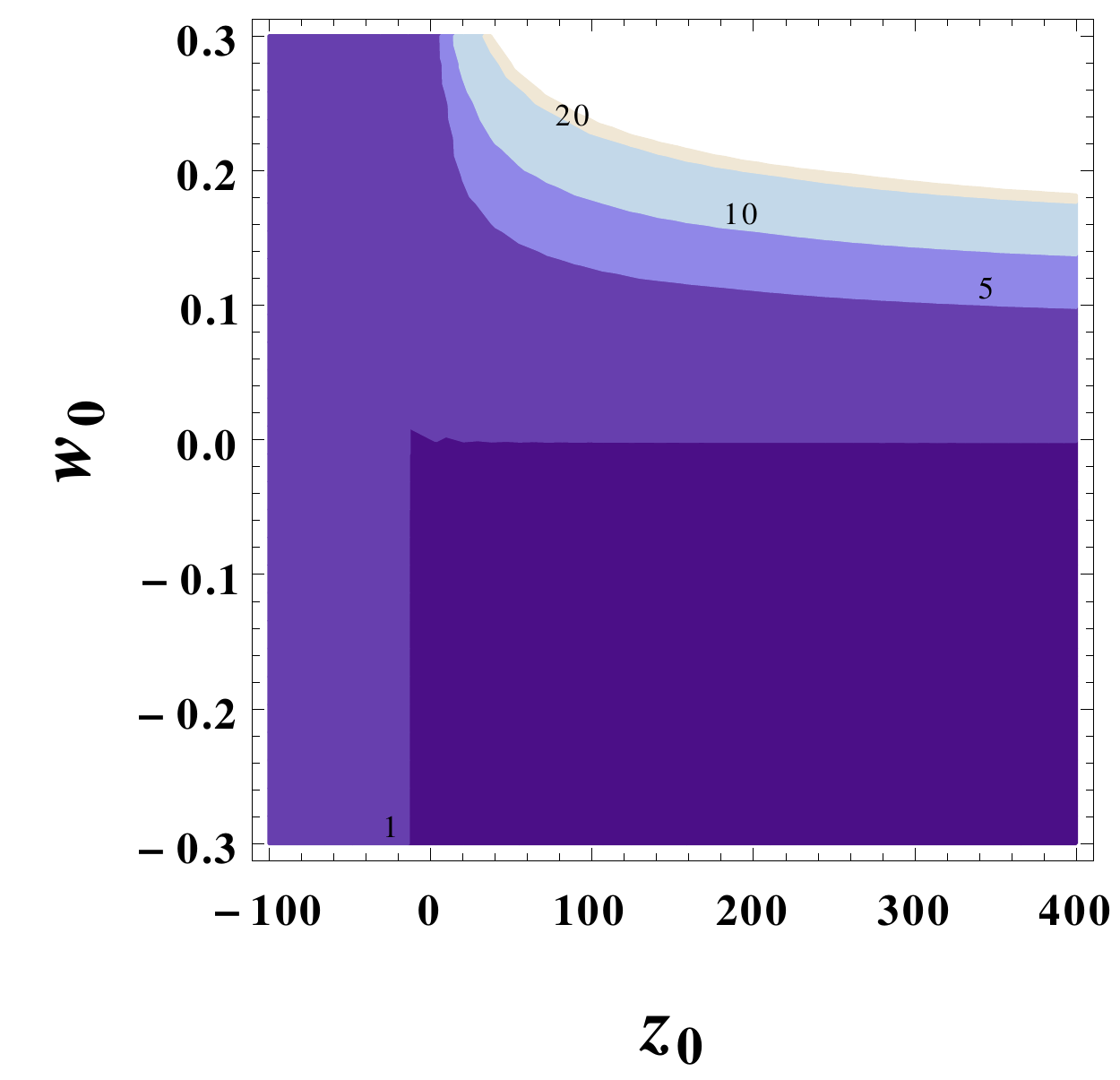}}
  \subfloat{\includegraphics[width=0.475\textwidth]{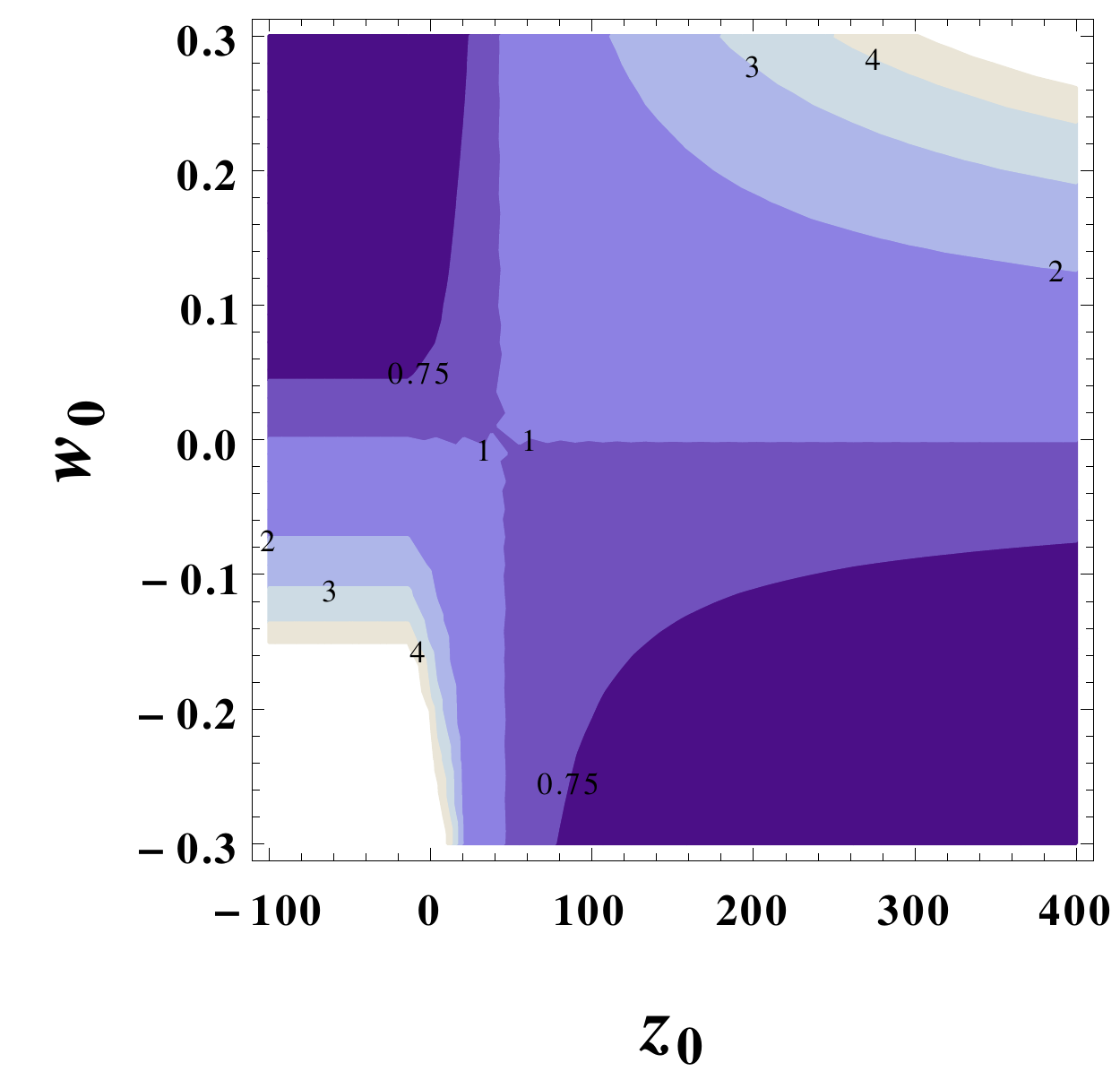}}
\caption{Values of $E_0=H(0)/H_0$ as a function of the parameters $w_0$ and 
$z_0$ for the EoS parameters (\ref{2.1}) and (\ref{2.2}).} \label{fig:3}
\end{figure}

\begin{figure}[h!]
  \centering
  \subfloat{\includegraphics[width=0.475\textwidth]{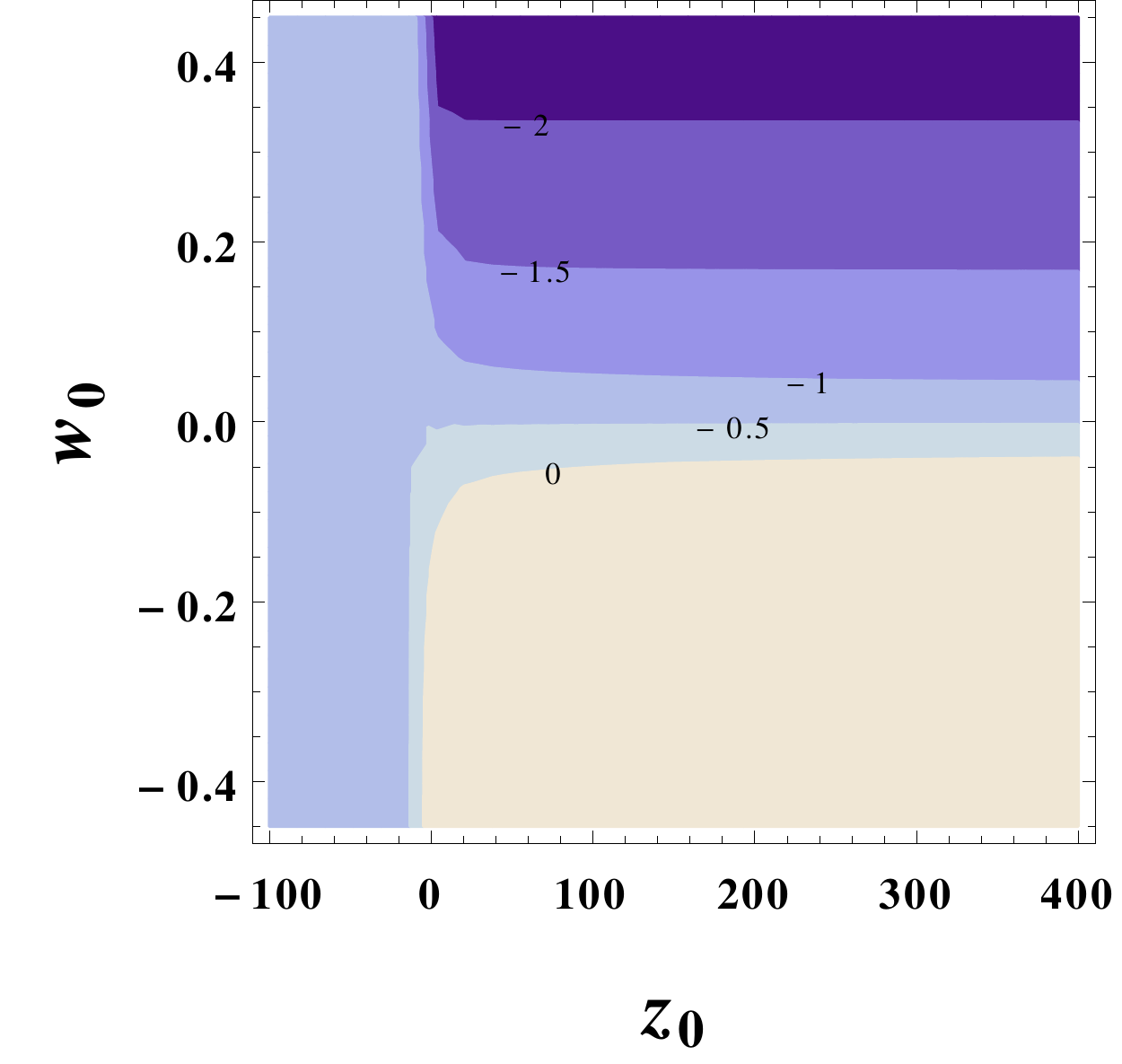}}
  \subfloat{\includegraphics[width=0.475\textwidth]{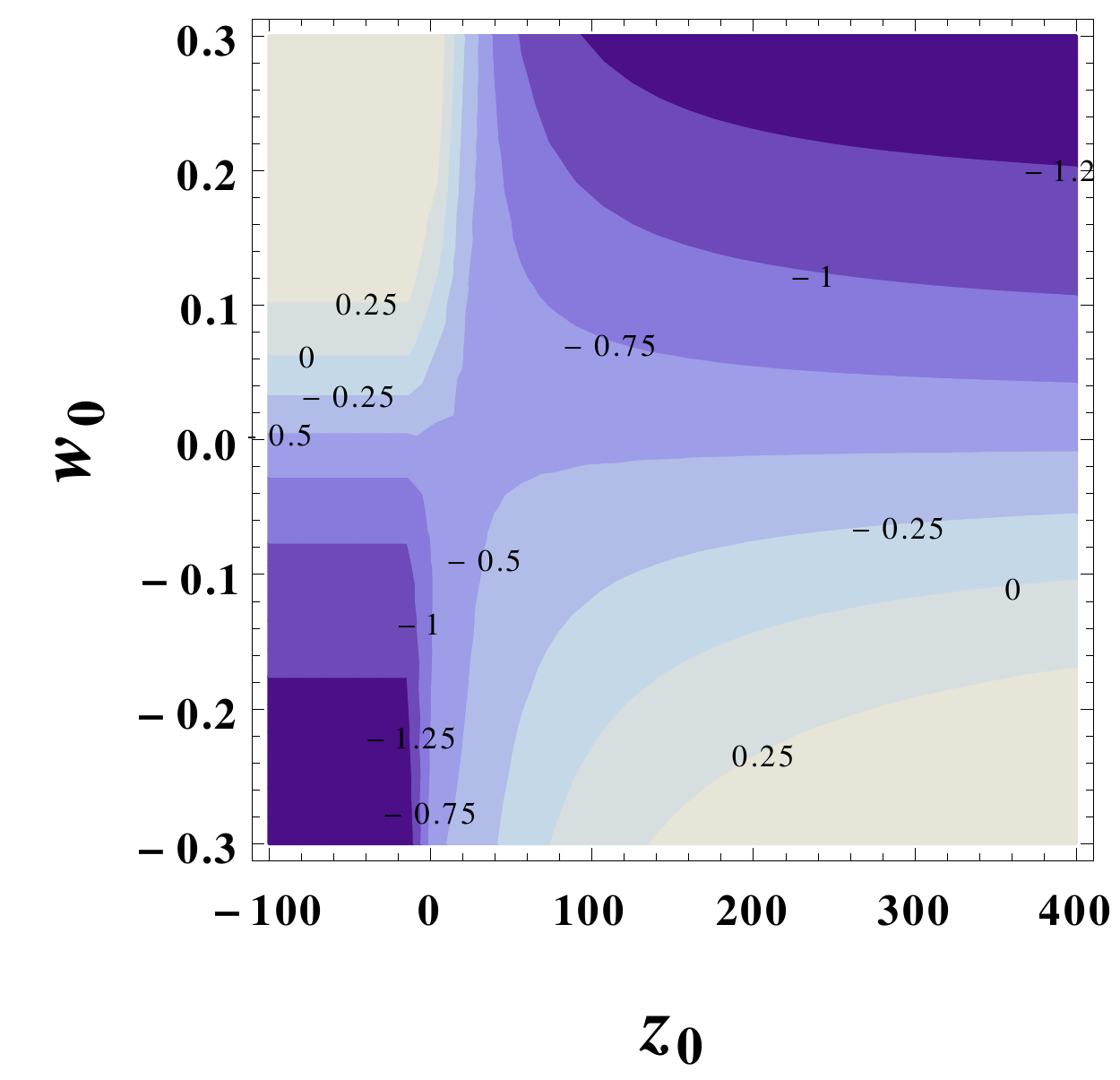}}
\caption{Values of $q_0$  as a function of the parameters $w_0$ and 
$z_0$ for the parametrizations (\ref{2.1}) and (\ref{2.2}).} \label{fig:4}
\end{figure}

%Then, as shown in Figs.~\ref{fig:1} and \ref{fig:2}, the cosmological evolution 
%followed by the parametrizations (\ref{2.1} and \ref{2.2}) fit well  the $\Lambda$CDM model at large 
%redshifts but the large differences can be easily appreciated at small 
%redshifts. 
In order to analyze these differences with more accuracy,  
Figs.~\ref{fig:3}-\ref{fig:4} show the values of $E(0)=H(0)/H_0$ and $q(0)$ evaluated 
today. Recalling that $E_{\Lambda CDM}(0)=1$ and $q_{\Lambda 
CDM}(0)=-0.5$ (with $\Omega_m^0=0.3$), the different predictions at $z=0$ can be easily compared. 

Hence, the cosmological evolution is well reproduced by the EoS given in 
(\ref{2.1}) and (\ref{2.2}), where the free parameters can be restricted to 
avoid a large deviation from $\Lambda$CDM. 
%In addition, in the following 
%section, the free parameters will be matched with the data coming 
%from SNe Ia observations as well as CMB and BAO. 
However, note that the above 
models, and in general models with an EoS parameter $w<-1$ may imply the occurrence of future 
singularities. The study of future singularities has drawn much attention over the last 
years, mainly because of under certain conditions some  realistic models with 
an appropriate  cosmological evolution that satisfy the observational 
constraints, may give rise to  some type of future 
singularity (see Ref.~\cite{phantom}), but also because of theoretical implications since possible quantum effects close to the singularity become important. A  classification of future singularities was proposed in Ref.~\cite{Nojiri:2005sx},

\begin{itemize}
\item Type I (``Big Rip''): For $t\rightarrow t_s$, $a\rightarrow
\infty$ and $\rho\rightarrow \infty$, $|p|\rightarrow \infty$.
\item Type II (``Sudden''): For $t\rightarrow t_s$, $a\rightarrow
a_s$ and $\rho\rightarrow \rho_s$, $|p|\rightarrow \infty$ (see Ref.~\cite{barrow1}).
\item Type III: For $t\rightarrow t_s$, $a\rightarrow a_s$
and $\rho\rightarrow \infty$, $|p|\rightarrow \infty$.
\item Type IV: For $t\rightarrow t_s$, $a\rightarrow a_s$ and
$\rho\rightarrow \rho_s$, $p \rightarrow p_s$
but higher derivatives of Hubble parameter diverge.
\end{itemize}

Nevertheless, the fact that the universe crosses the phantom barrier  is not a sufficient condition for the occurrence of a future singularity, and may lead to other 
kind of non singular scenarios \cite{LittleRip}. For a non constant EoS parameter $w$, the presence of future 
singularities depends on the asymptotic behavior of the EoS parameter. Then, by assuming a large value of the scale factor in comparison with the one today, $a \ggg 1$, when 
pressureless matter becomes negligible,  the above models (\ref{2.1}) and (\ref{2.2}) can be approximated as follows
\bea
w_1(a)=-1+ w_0\left[ \tanh\left(\frac{1}{a} -1 -z_0\right)-1\right]\sim -1- w_0\left[ \tanh\left(1+z_0\right)+1\right]=\tilde{w_1} \ , \nn
w_2(a)= -1+ w_0\tanh\left(\frac{1}{a} -1 -z_0\right)\sim -1-w_0 \tanh\left(1+z_0\right) = \tilde{w_2}\ . 
\label{s.2}
\eea
Hence, the EoS parameters (\ref{2.1}) and (\ref{2.2}) become constant  in the far future, whose value depends on the parameters $w_0$ and $z_0$. This coincides with the above analysis regarding the potential and the kinetic term of the scalar field that reproduces such models, where was found that the scalar field leads to a constant EoS asymptotically. Then, it is straightforward  to solve  the FLRW equations (\ref{ST2}) which yield 
\be
H\sim\frac{2}{3|1+\tilde{w}|}\frac{1}{t_s-t}\ ,
\label{s.2a}
\ee
where $1+ \tilde{w} < 0$ has been assumed and $t_s$ is the so called Rip time, the remained time  for the occurrence of the
Big Rip singularity. 
%Fig.~\ref{fig:s.1} depicts  the range of values of 
%the free parameters for which the models are free of future singularities. \\
%\begin{figure}[h]
%  \centering
%  \subfloat{\includegraphics[width=0.475\textwidth]{images/Sing-w1}}
%  \subfloat{\includegraphics[width=0.475\textwidth]{images/Sing-w2}}
%\caption{Values of $\tilde{w}$ for $w_1$ (left) and $w_2$ (right) for different 
%values of their free parameters. White area corresponds to a model free of 
%future singularities ($\tilde{w} > -1$), purple corresponds to a universe where 
%a Big Rip singularity occurs ($\tilde{w} < -1$).} \label{fig:s.1}
%\end{figure}
Hence, a future singularity will occur in case that $w_0>0$ in (\ref{2.1}), whereas  the expansion would evolute smoothly for  $w_0<0$. Moreover, $\tilde{w}_1$ approaches $-1$
at low redshifts  in case that $z_0$ takes negatives values independently of $w_0$  leading effectively to 
$\Lambda$CDM. In the second parametrization, the value of $\tilde{w}_2$ is greater than -1 for negative 
(positive) values of $w_0$ and $z_0 > -1$ ($z_0 < -1$), so there is no 
singularity. For positive (negative) values of $w_0$ and $z_0 > -1$ ($z_0 < 
-1$), the value of $\tilde{w}_2$ is below $-1$ and a singularity emerges, whereas 
$\tilde{w}_2$ tends asymptotically to $-1$ for $w_0=0$ and/or $z_0=-1$.  \\

Thus, depending on the free parameters, the above models may lead to some kind of future singularity.\\

%%%%%%%%%%%%%%%%%%%%%%%%%%%%%%%%%%

\section{Fitting the models with SNe Ia data and standard rulers}
\label{ObsData}

%%%%%%%%%%%%%%%%%%%%%%%%%%%%%%%%%

Firstly we compute the best fit for the above parametrizations by using the SNe dataset of $557$ SN stars of union2 (see 
Ref.~\cite{Suzuki:2011hu}). Here we use the technique of the maximum likelihood to 
find the best fit of the parameters (see for instance \cite{Lazkoz:2005sp}). Then, for a 
particular set of the free parameters, the Hubble parameter 
$H(z;\Omega_m^0,w_0,z_0)$ can be computed by solving the equation (\ref{ST7}), and the corresponding Hubble free luminosity distance is obtained,
\begin{equation}
D_L^{th} (z;\Omega_m^0, w_0 ,z_0)= (1+z) \int_0^z dz'\frac{H_0}{H(z';\Omega_m^0, w_0 ,z_0)}\ . 
\label{SN1} 
\end{equation}
Whereas the apparent magnitude is connected to the free luminosity distance by 
\begin{equation}
m(z;\Omega_m^0, w_0,z_0)={\bar M} (M,H_0) + 5 log_{10} (D_L (z;\Omega_m^0, w_0 ,z_0))\ , 
\label{SN2} 
\end{equation}
where ${\bar M}$ is the magnitude zero point offset and depends on the absolute 
magnitude $M$ and on the present Hubble parameter $H_0$ as 
\begin{equation}
{\bar M} = M + 5 log_{10}(\frac{c\; H_0^{-1}}{Mpc}) + 25\ . 
\label{SN3} 
\end{equation}
Hence by using the observational data  from \cite{Suzuki:2011hu}, where the apparent magnitudes $m(z)$ of the SN Ia with 
the corresponding redshifts $z$ and errors $\sigma_{m(z)}$ are obtained, the best fit 
corresponding to our parameters $\{\Omega_m^0, w_0, z_0\}$ is determined by the probability 
distribution
\begin{equation}
P({\bar M}, \Omega_m^0, w_0, z_0)= {\cal N} e^{- \chi^2({\bar M}, \Omega_m^0, w_0, z_0)/2}\ , 
\label{SN4} 
\end{equation} 
where  
\begin{equation}
\chi^2 ({\bar M}, \Omega_m^0, w_0, z_0)= \sum_{i=1}^{557} \frac{(m^{obs}(z_i) - m^{th}(z_i;{\bar M}, \Omega_m^0, w_0, z_0))^2} {\sigma_{m^{obs}(z_i)}^2}\ , 
\label{SN5} 
\end{equation}
and ${\cal N}$ is a normalization factor. The parameters $\{{{\bar \Omega}_m^0, \bar 
w}_0,{\bar z}_0\}$ that minimize the $\chi^2$ expression (\ref{SN5}) are the `best 
fit' and the corresponding $\chi^2({\bar \Omega}_m^0, {\bar w}_0,{\bar z}_0)\equiv \chi_{min}^2$ 
gives an indication of the quality of the particular parametrization: the smaller 
$\chi_{min}^2$ is, the better the parametrization.

We can trivially minimize the parameter $\bar{M}$ by expanding the $\chi^2$ in 
equation (\ref{SN5}) with respect to $\bar{M}$ as 
\begin{equation}  
\chi^2 (\Omega_m^0, w_0, z_0)= A - 2 {\bar M} B  + {\bar M}^2 C\ ,  
\label{SN6} 
\end{equation}
where 
\begin{eqnarray}
A(\Omega_m^0, w_0, z_0)&=&\sum_{i=1}^{557} \frac{(m^{obs}(z_i) - m^{th}(z_i ;{\bar M}=0, \Omega_m^0, w_0, z_0))^2}{\sigma_{m^{obs} (z_i)}^2} 
\label{SN7.1} \nonumber \\
B(\Omega_m^0, w_0, z_0)&=&\sum_{i=1}^{557} \frac{(m^{obs}(z_i) - m^{th}(z_i ;{\bar M}=0, \Omega_m^0, w_0, z_0))}{\sigma_{m^{obs}(z_i)}^2} 
\label{SN7.2} \nonumber \\
C&=&\sum_{i=1}^{557}\frac{1}{\sigma_{m^{obs}(z_i)}^2 } 
\label{SN7.3}
\end{eqnarray} 
Then, equation (\ref{SN6}) has a minimum at ${\bar M}={B}/{C}$ given by 
\begin{equation}
{\tilde\chi}^2(\Omega_m^0, w_0, z_0)=A(\Omega_m^0, w_0, z_0)- \frac{B(\Omega_m^0, w_0, z_0)^2}{C} 
\label{SN8}
\end{equation}
Hence, instead of minimizing $\chi^2({\bar M}, \Omega_m^0, w_0, z_0)$ we can minimize 
${\tilde\chi}^2(\Omega_m^0, w_0, z_0)$ independently of ${\bar M}$. Obviously 
$\chi_{min}^2={\tilde\chi}_{min}^2$ and in what follows the tilde is omitted for 
simplicity. Furthermore, the reduced $\chi^2_{red}$ is also computed in order to compare both models with $\Lambda$CDM.\\

\begin{figure}[h]
  \centering
  \subfloat{\includegraphics[width=0.475\textwidth]{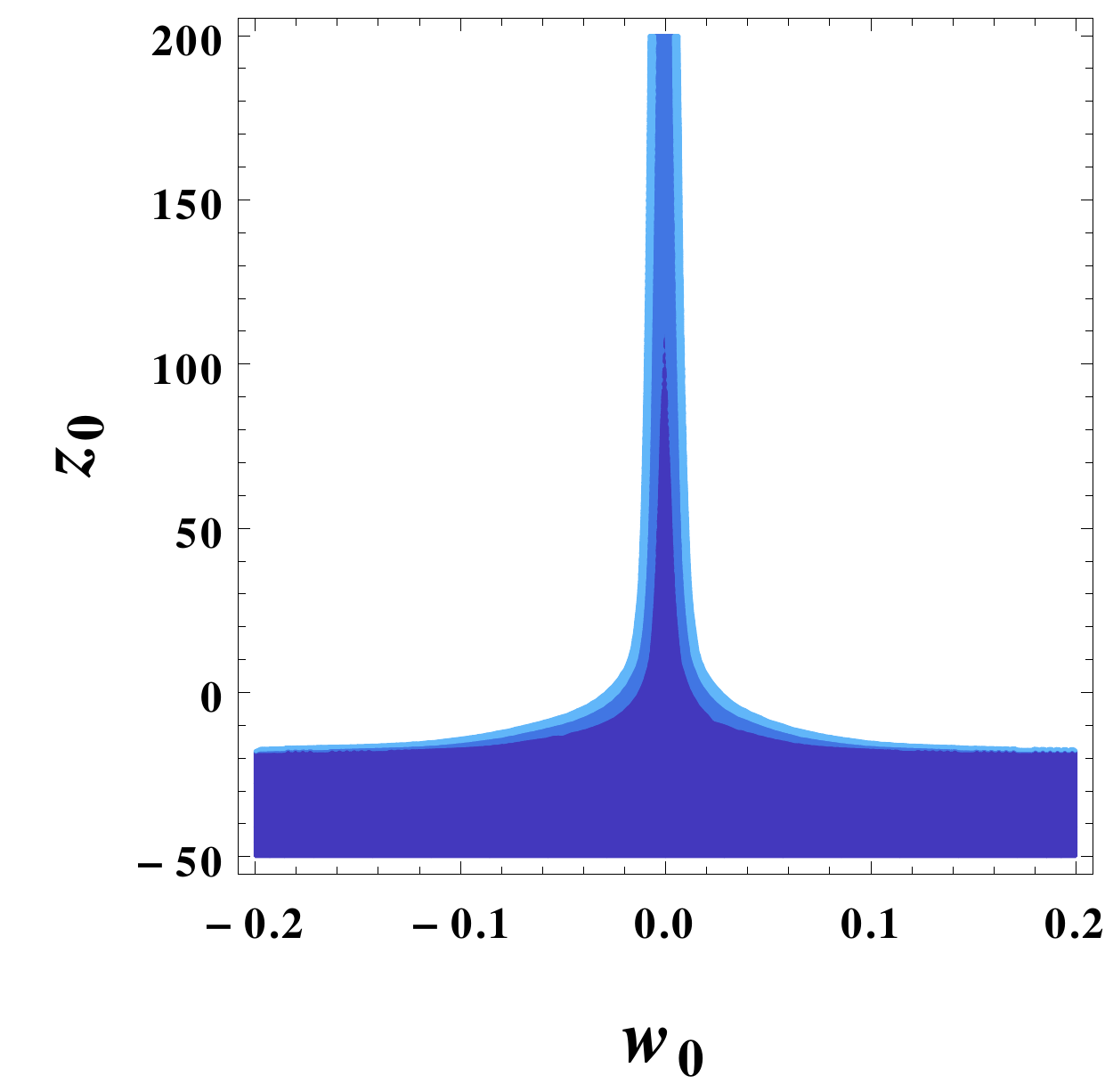}}
  \subfloat{\includegraphics[width=0.475\textwidth]{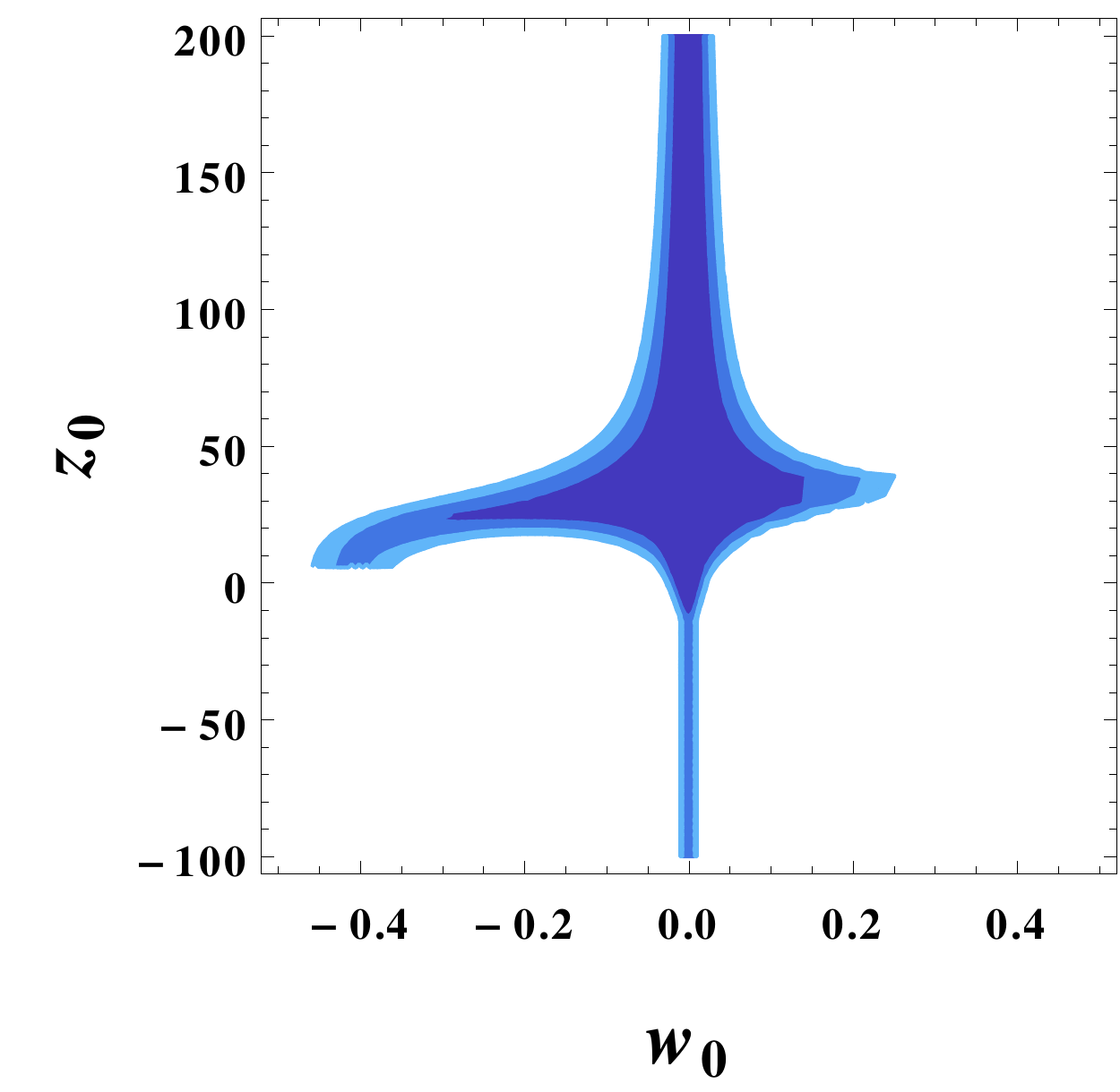}}
\caption{Contour plots for the parameters $w_0$ and $z_0$ for the first (left) and second (right) models taking $\Omega_{0m}=0.27$.} \label{fig:SN.1}
\end{figure}

Let us consider an initial computation by assuming  the best fit $\Omega_m^0=0.27$ for the $\Lambda$CDM model. Then, $\chi^2=\chi^2(w_0, z_0)$ and the results are shown in Table \ref{table1} and Fig.~\ref{fig:SN.1}, where the $\chi_{min}^2$ value  have a similar value in both parametrizations in comparison with  
the $\Lambda$CDM model and even lower for the parametrization $w_1$. the reduced $\chi^2$, defined as
\be
\chi^2_{red} =\dfrac{\chi^2_{min}}{N_{data}-dof-1}
\ee
where $N_{data}$ is the number of experimental points used and the degree of freedom $dof$ is the number of parameters of the model, shows a better fit for the $\Lambda$CDM model. In both cases, $w_0=0$ corresponds to a cosmological constant, but does not coincide with the best fit, which is slightly displaced from that point whereas the $z_0$ parameter presents a large error since the second part of both parametrizations does not contribute when setting $w_0=0$. In 
both cases there are possibilities for the occurrence of a Big Rip singularity within the confidence region.\\
%\begin{widetext}
\vspace{-2pt}
\begin{table}[h!]  %%%%%%%%% Table of EOS parametrizations %%%%%%%%%%%
\begin{center}
\begin{minipage}{0.88\textwidth}
\caption{Best fit for the models (\ref{2.1}) and (\ref{2.2}) with $\Omega_{0m}=0.27$ by using the Sne Ia dataset \cite{Suzuki:2011hu}. The result for the $\Lambda$CDM model is also shown. \label{table1}}
\end{minipage}\\
\begin{tabular}{ccccccc}
\hline
\hline
\bf{Model} & $\bf{\chi_{min}^2}$ &  $\bf{w_0}$ & $\bf{z_0}$ & $\bf{\Omega_{0m}}$ & $\bf{\chi_{red}^2}$\\
\hline \vspace{-5pt}\\
$\Lambda$CDM  & $542.685$ & - & - & $0.27 \pm 0.02$ & $0.978 $\\
\\
$ w_1(z)$ & $542.683$ & $0.0045 \pm 0.1$ & $-25 \pm 30$ & $0.27$ & $0.981 $ \\
\\
$ w_2(z)$ & $541.583$ & $-0.03 \pm 0.07$ & $22 \pm 45$ & $0.27$ & $0.979 $\\\\
 \hline \hline
\end{tabular}
\end{center}
\end{table}
%\end{widetext}

\begin{figure}[h]
  \centering
  \subfloat{\includegraphics[width=0.475\textwidth]{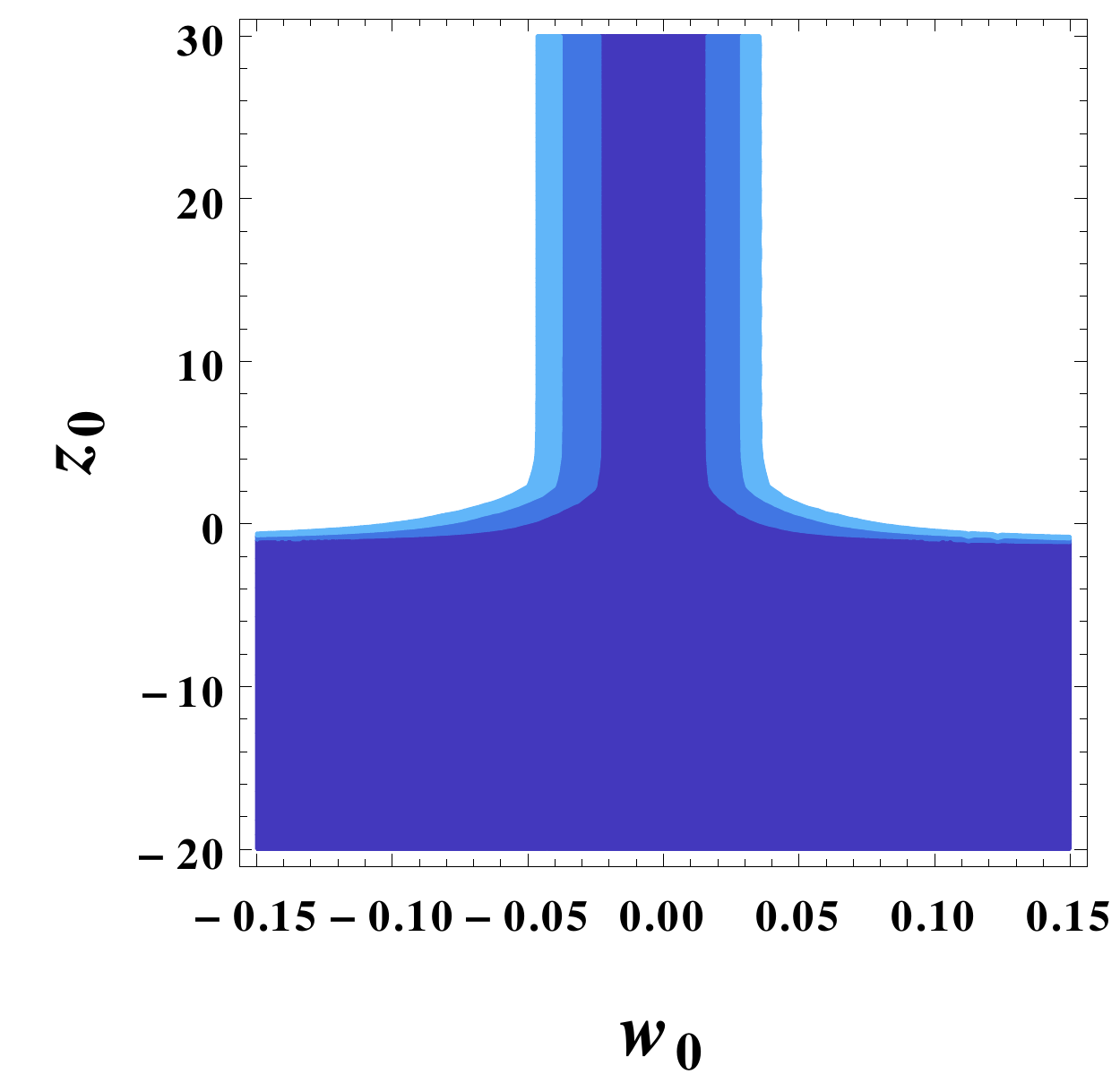}}
  \subfloat{\includegraphics[width=0.475\textwidth]{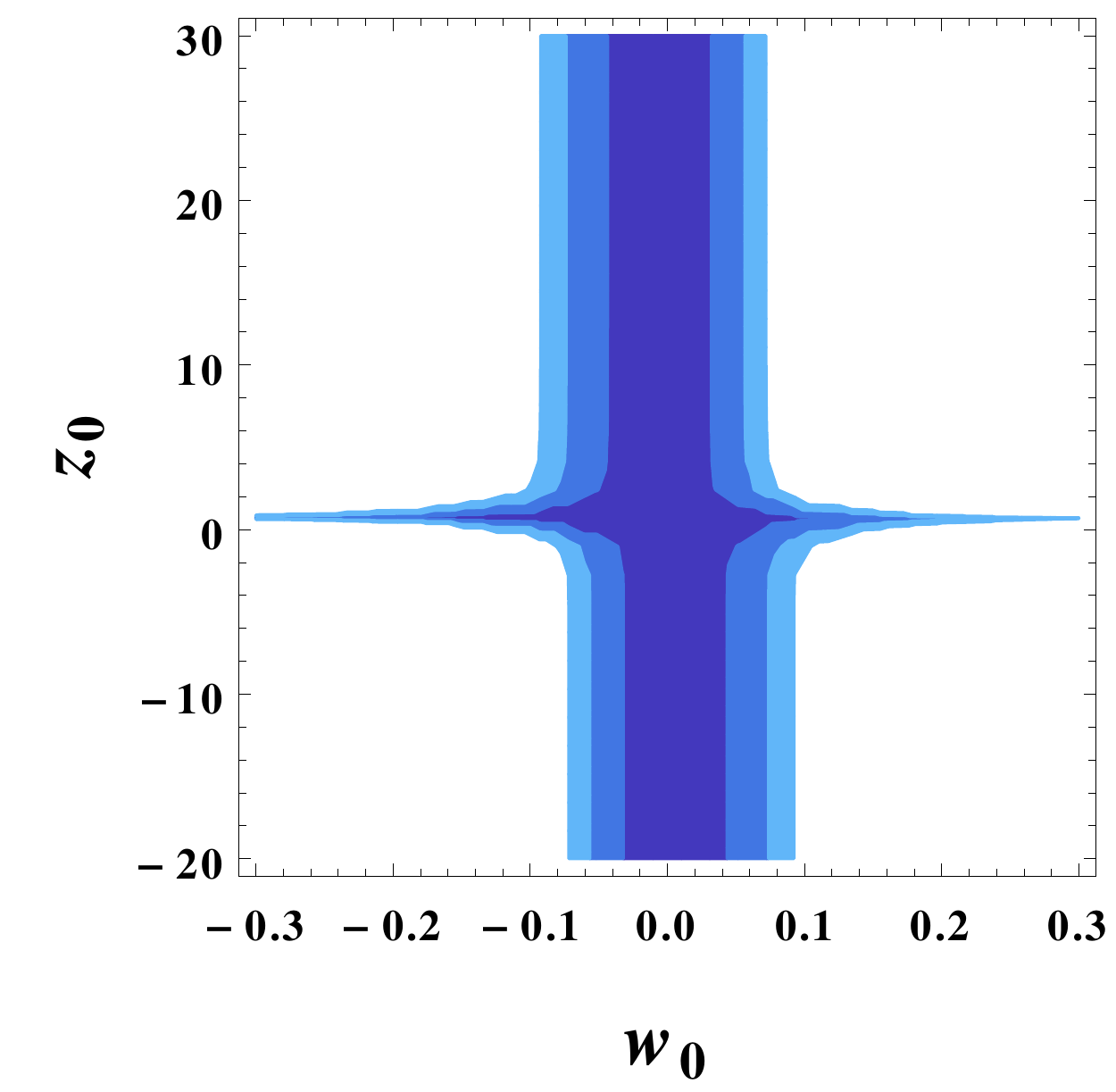}}
\caption{Contour plots of the parameters $w_0$ and $z_0$ for $w_1$ (left) and $w_2$ (right) models using Standard Rulers data.} \label{fig:SR.1}
\end{figure}
\begin{figure}[Hh!]
\begin{minipage}{1.0\textwidth}
 \centering
  \includegraphics[width=0.475\textwidth]{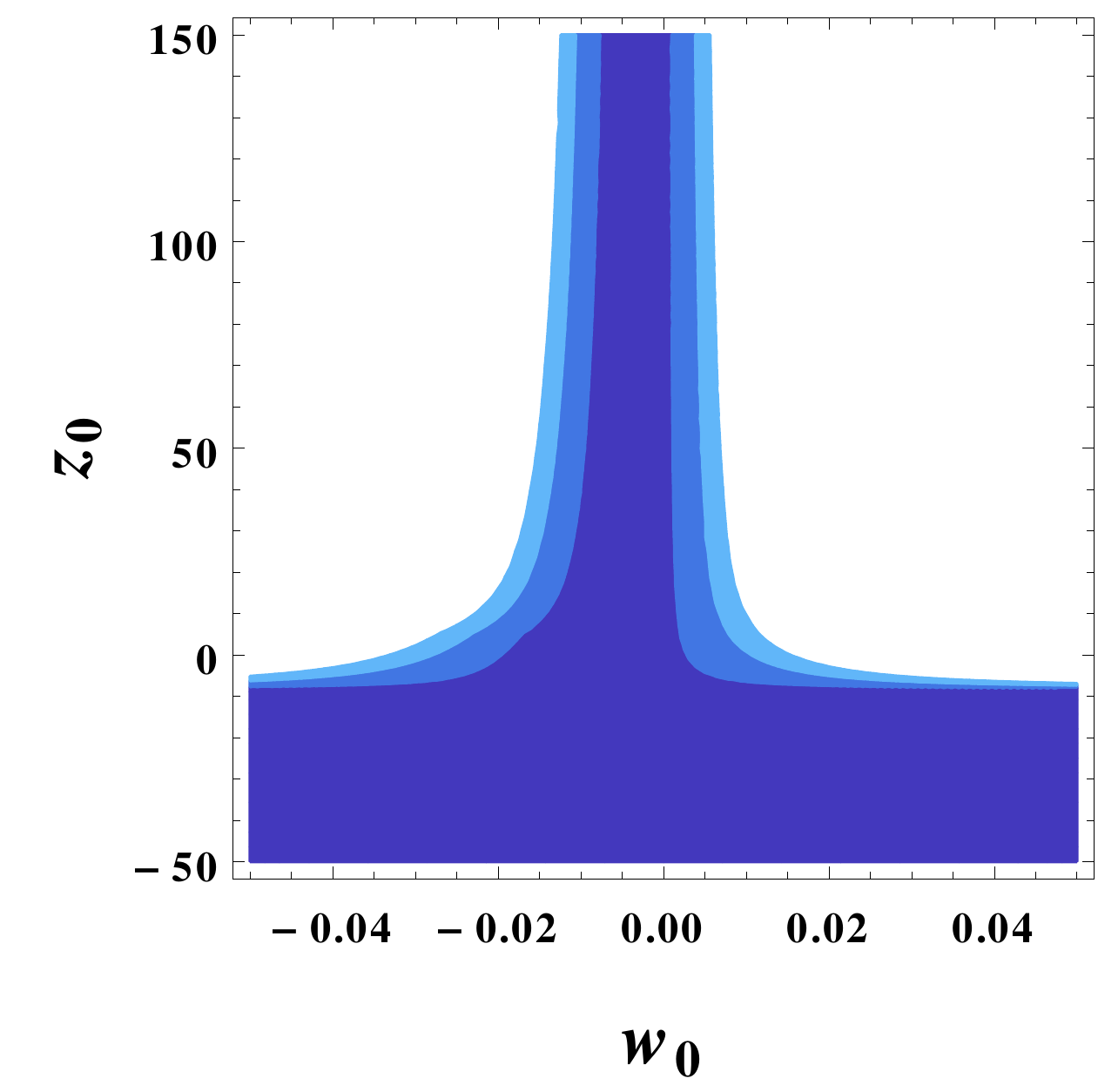}
  \includegraphics[width=0.5\textwidth]{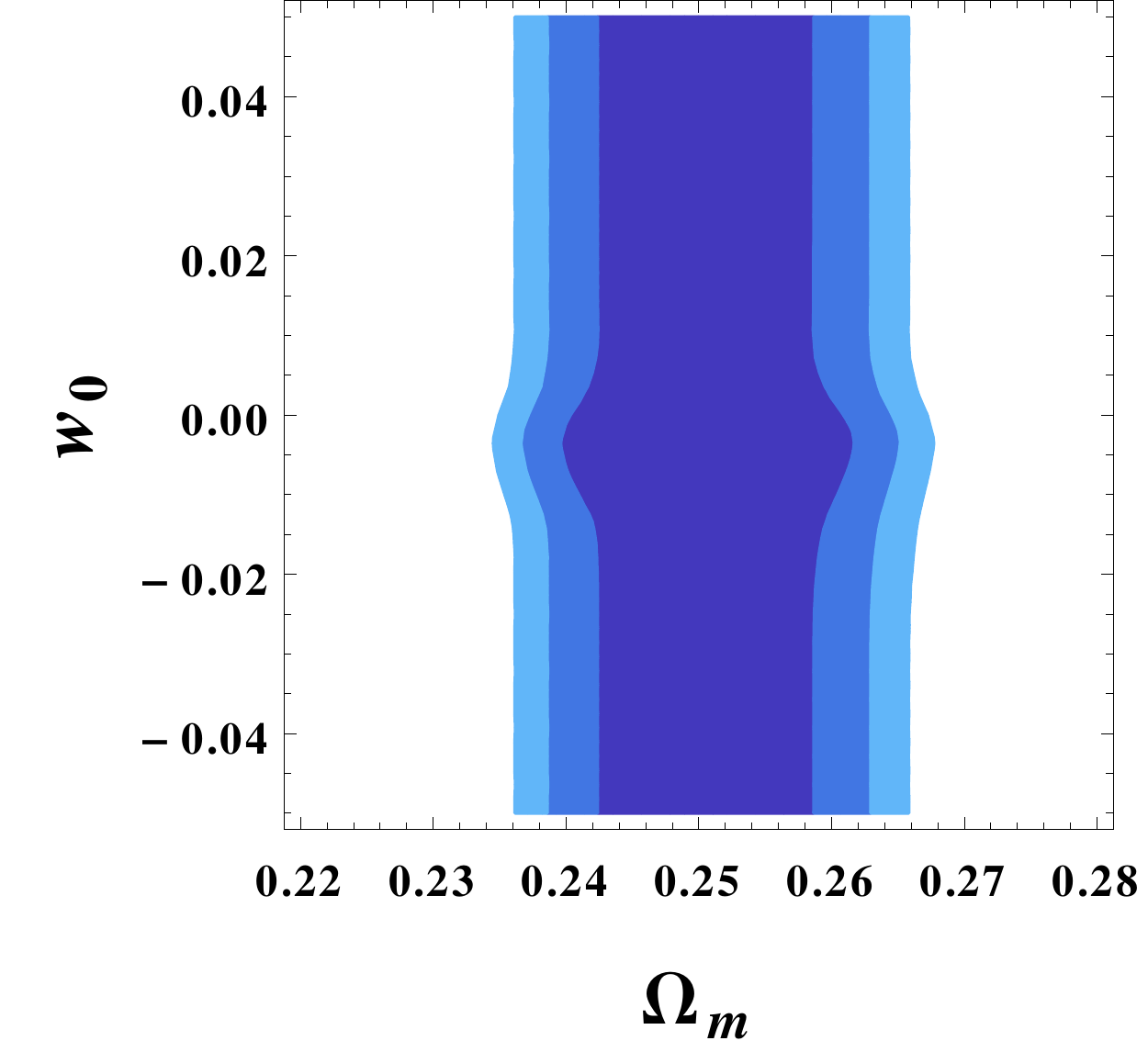}
\end{minipage}  
 \centering
  \includegraphics[width=0.475\textwidth]{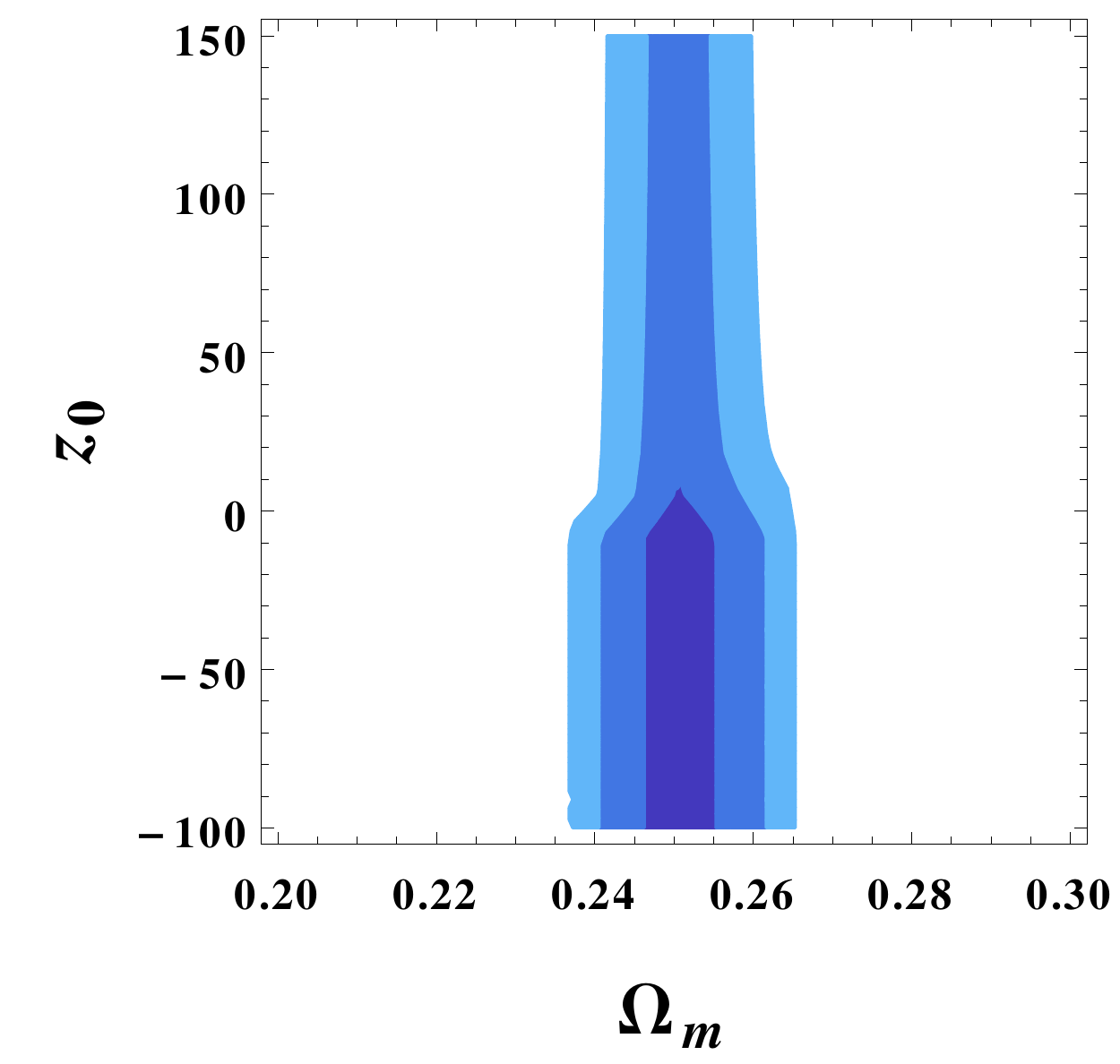}
\caption{Contour plots for the first model (\ref{2.1}) when using Sne Ia data and Standard Rulers.} \label{fig:C.1}
\end{figure}
\begin{figure}[Hh!]
\begin{minipage}{1.0\textwidth}
 \centering
  \includegraphics[width=0.475\textwidth]{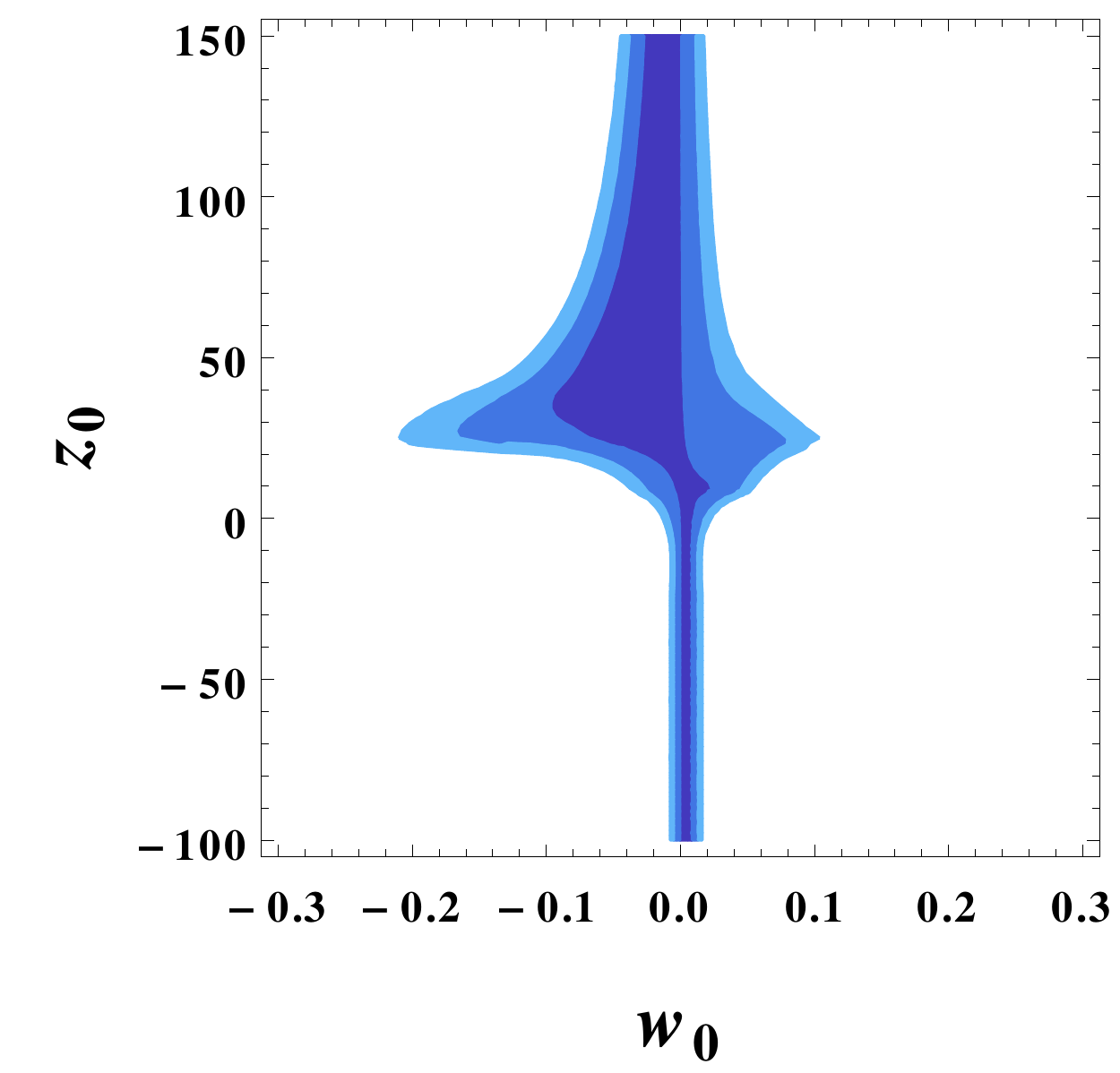}
  \includegraphics[width=0.475\textwidth]{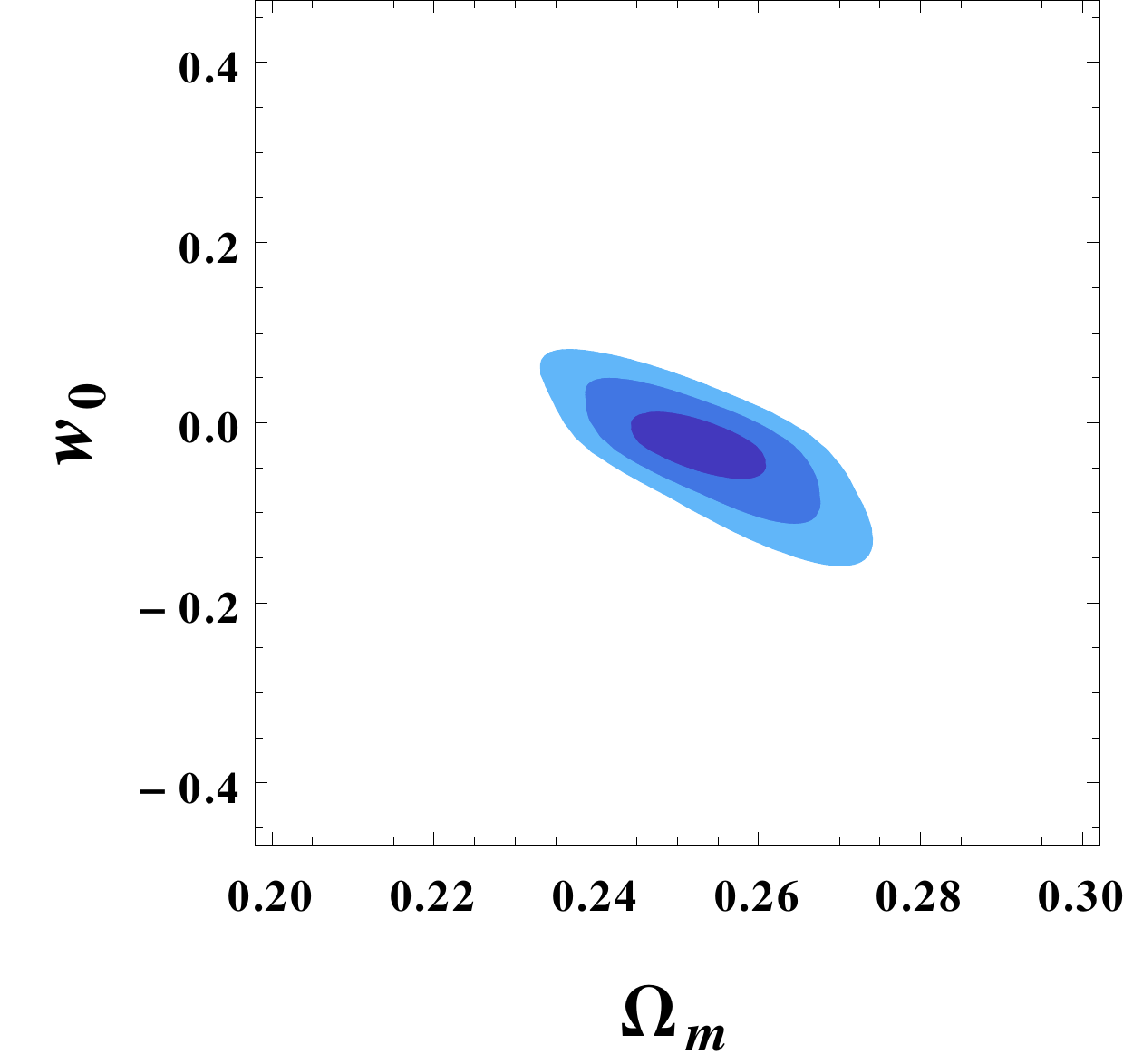}
\end{minipage}  
 \centering
  \includegraphics[width=0.475\textwidth]{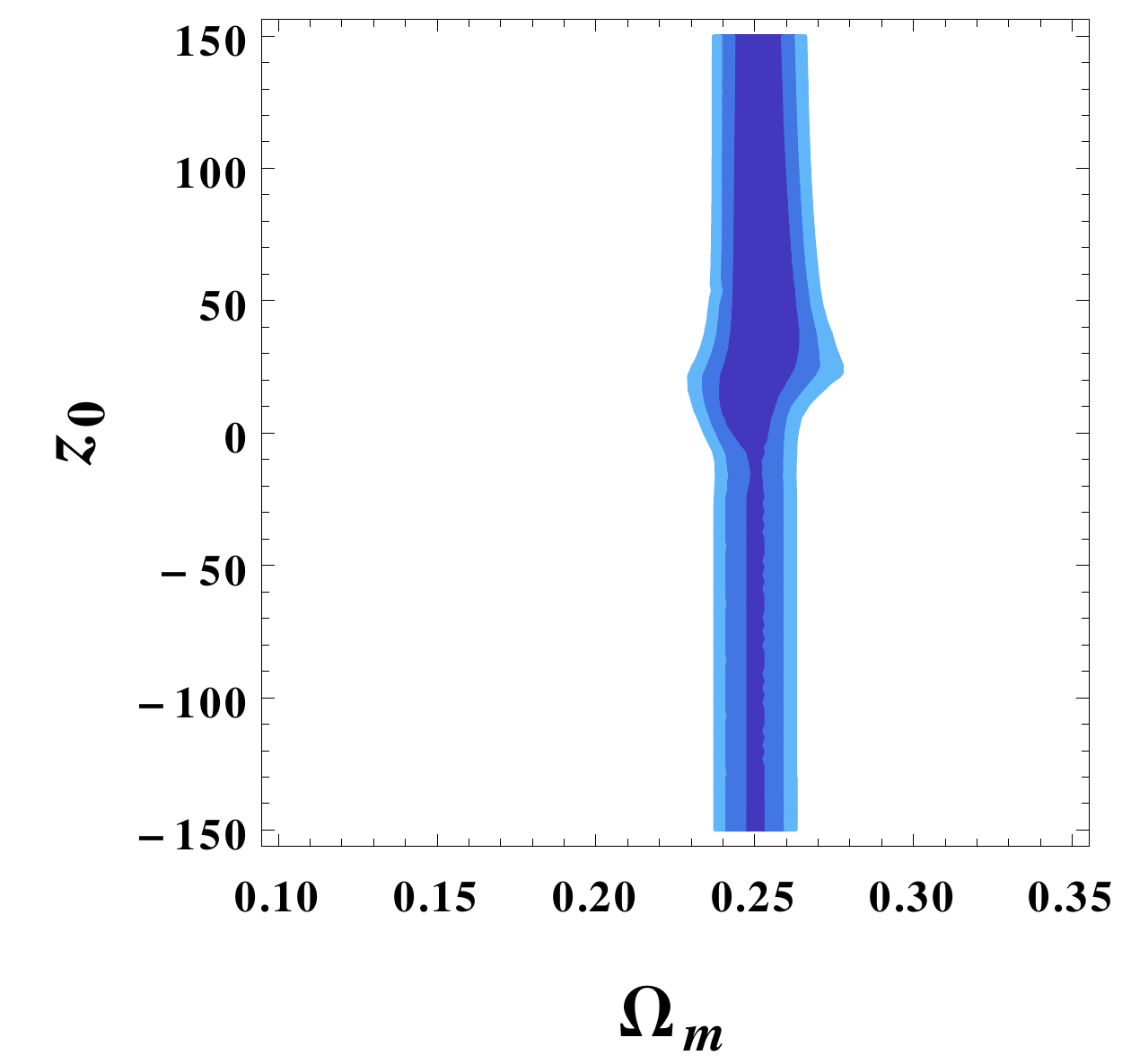}
\caption{Contour plots for the second model (\ref{2.2}) when combining Sne Ia data and Standard Rulers.} \label{fig:C.2}
\end{figure}

%%%%%%%%%%%%%%%%%%%%%%%%%%%%%%%%%%%
%
%\section{Fitting the models with Standard Ruler}
%\label{SR}
%
%%%%%%%%%%%%%%%%%%%%%%%%%%%%%%%%%%

Let us now use the Standard Ruler data to test
the models (\ref{2.1}) and (\ref{2.2}). Standard Rulers are objects of 
known comoving size which may be used to measure the angular diameter distance. 
This data comes from two different sources: the Cosmic Microwave Background (CMB) \cite{Spergel:2006hy} and  the Baryon Acoustic 
Oscillations (BAO) \cite{Eisenstein:2005su}.\\

In order to compute the theoretical points at early times it is necessary to 
consider the radiation contribution, $p_{rad}=\frac{1}{3}\rho_{rad}$, so that the FLRW equations (\ref{ST2}) yield
\be
H^{2} = \frac{\kappa^2}{3} \left( \rho_{m} + \rho_{rad} + \rho_{_{\phi}} \right)\ ,\quad
\dot{H} = - \frac{\kappa^2}{2} \left[ \rho_{m} + \frac{4}{3} \rho_{rad} + (1+w_{\phi})\rho_{_{\phi}} \right]\ . 
\label{EFeq}
\ee
By using the expressions of 
relative densities $\Omega_i^{0}=\rho_{0i}/\rho_c$, the FLRW 
equations (\ref{EFeq}) are describe as
\bea
E^{2} = \left[ \Omega_{m}^0\ a^{-3} + \Omega_{rad}^0\ a^{-4} + \Omega_{\phi}^0 
X(a) \right]\ \\
\dot{E} = - \frac{3}{2} \left[ \Omega_{m}^0\ a^{-3} + 
\frac{4}{3}\Omega_{rad}^0\ a^{-4} + (1+w_{\phi})\Omega_{\phi}^0 X(a) \right] \ , 
\label{EFeq2}
\eea
where $E=H/H_0$, $\Omega_{m}^0 + \Omega_{rad} + \Omega_{\phi}=1$ for a flat 
universe and $X(a)$ is defined in terms of the scale factor as follows
\be
X(a)={\rm exp}\left[-3\int_{1}^a\! \frac{(1+w(a'))}{a'}da'\right]\ .
\label{--}
\ee
Since $\Omega_{m}^0$ and $\Omega_{rad}^0$ can be related, the above equations may be rewritten as follows
\bea
E^2(a)=\Omega_m(a+a_{eq})a^{-4}+ \Omega_{de}{X(a)} \\
\dot{E} = - \frac{3}{2} \left[ \Omega_{m}^0(a + \frac{4}{3} a_{eq}) a^{-3} + (1+w_{DE})\Omega_{DE}^0 X(a) \right] \ , 
\label{EFeq3}
\eea
where $a_{eq}=\Omega_{rad}/\Omega_m$, which can be  expressed in 
terms of the redshift $a_{eq}=1/(1+z_{eq})$, where the equilibrium redshift is defined as the redshift when matter (baryons, electrons, and CDM) and radiation (photons and massless neutrinos) had the same density, $z_{eq}=2.5\times 10^4 \Omega_m h^2 (T_{CMB}/2.7\,{\rm K})^{-4}$, being 
$T_{CMB}$ the photon temperature of the CMB (see \cite{Eisenstein:1997ik}).
\\

%Thus, we take the first equation of (\ref{EFeq3}) :
%\be 
%E^2(a)\equiv \frac{H(a)^2}{H_0^2}=\Omega_m(a+a_{eq})a^{-4}+ \Omega_{de}{X(a)} \ ,
%\label{SR1}
%\ee
%where $a_{eq}=\Omega_{rad}/\Omega_m=1/(1+z_{eq})$, and we compute its value by 
%using $z_{eq}=2.5\times 10^4 \Omega_m h^2 (T_{CMB}/2.7\,{\rm K})^{-4}$ and 
%taking $T_{CMB}=2.728$K \cite{Wang:2007mza}.\\

The $\chi^2_{CMB}$ is computed by using the dataset $(R,l_a,\Omega_b 
h)$, \cite{Spergel:2006hy}, where the first point is the scaled distance to 
recombination given by
\be 
R=\sqrt{\Omega_{0m} \frac{H_0^2}{c^2}} \; r(z_{CMB})\ ,
\label{R}
\ee
where $r(z_{CMB})$ is the comoving distance,
\be
r(z)=\frac{c}{H_0} \int_0^z \frac{dz}{E(z)}\ .
\label{r}
\ee
The second point corresponds to the angular scale of the sound horizon at recombination,
\be
l_a=\pi \frac{r_(a_{CMB})}{r_s(a_{CMB})}\ ,
\label{la}
\ee
where $a_{CMB}=\frac{1}{1+z_{CMB}}$ with $z_{CMB}=1089$ and being 
$r_s(a_{CMB})$ the comoving sound horizon at recombination,
\be
r_s(a_{CMB})=\frac{c}{H_0} \int_0^{a_{CMB}} \frac{c_s(a)}{a^2 E(a)} da\ ,
\label{rs}
\ee
with the speed of sound $c_s(a)=1/\sqrt{3(1 + \bar{R}_b a )}$, $\bar{R}_b=\frac{3}{4} \frac{\Omega_b h^2} {\Omega_{\gamma} h^2}= 31500 \Omega_b h^2 (T_{CMB}/2.7 K)^{-4}$ being the photon-baryon energy-density ratio and $\Omega_b$ the baryon density. The observational data is given by \cite{Spergel:2006hy}
\begin{eqnarray}
\bf{{\bar V}_{CMB}} &=& \left( \begin{array}{c}
{\bar R} \\
{\bar l_a}\\
{\bar \Omega_b h}\end{array} \right)= \left(\begin{array}{c}
1.70\pm 0.03 \\
302.2 \pm 1.2\\
0.022 \pm 0.00082 \end{array} \right)
\label{Vcmb}
\end{eqnarray}
Whereas the inverse covariance matrix is 
\begin{eqnarray}
 {\bf C_{CMB}}^{-1}=\left(
\begin{array}{ccc}
 1131.32 & 4.8061 & 5234.42\\
 4.8061 & 1.1678 & 1077.22 \\
 5234.42 & 1077.22 & 2.48145 \times 10^6
\end{array} \right) .
\end{eqnarray}
Then, as usual the $\chi^2$ can be constructed by the difference between the 
experimental and theoretical points
\begin{eqnarray}
\bf{X_{CMB}} &=& \left(\begin{array}{c}
R - 1.70 \\
l_a -302.2\\
\Omega_b h^2 - 0.022 \end{array} \right),
\end{eqnarray}
and the contribution to the $\chi^2$ by using the CMB dataset yields
\be
\chi^2_{CMB}=\bf{X_{CMB}}^{T}{\bf C_{CMB}}^{-1}\bf{X_{CMB}} \ .
\label{chiCMB}
\ee 

On the other hand, Table~\ref{table3} contains the experimental points coming from the analysis of BAO, whereas the  inverse covariant matrix is given by \cite{Eisenstein:2005su}
\begin{eqnarray}
 {\bf C_{BAO}}^{-1}=\left(
\begin{array}{cccccc}
 4444.44 & 0 & 0 & 0 & 0 & 0\\
 0 & 30317 & -17312 & 0 & 0 & 0\\
 0 & -17312 & 87046 & 0 & 0 & 0\\
 0 & 0 & 0 & 1040.3 & -807.5 & 336.8\\
 0 & 0 & 0 & -807.5 & 3720.3 & -1551.9\\
 0 & 0 & 0 & 336.8 & -1551.9 & 2914.9
\end{array} \right).
\end{eqnarray}

%\begin{widetext}
\vspace{-2pt}
\begin{table}[h!]  %%%%%%%%% 6 data points for BAO %%%%%%%%%%%
\begin{center}
\begin{minipage}{0.88\textwidth}
\caption{Experimental BAO points: here $A(z)$ is the acoustic parameter  and 
$d_z=r_s(z_d)/D_V (z)$, being $D_V(z)$ is the dilation scale, Ref.~\cite{Eisenstein:2005su}. Points used for parameter 
fitting are bold font. \label{table3}}
\end{minipage}\\
\begin{tabular}{ccccccc}
\hline
\hline
\bf{Sample} & & $\bf{z}$ & & $\bf{d_z}$ & & $\bf{A(z)}$ \\
\hline %\vspace{-1pt}\\
6dFGS & & $0.106$ & & $\bf{0.336 \pm 0.015}$ & & $0.526 \pm 0.028$
\\
SDSS & & $0.2$ & & $\bf{0.1905 \pm 0.0061}$ & & $0.488 \pm 0.016$
\\
SDSS & & $0.35$ & & $\bf{0.1097 \pm 0.0036}$ & & $0.484 \pm 0.016$
\\
WiggleZ & & $0.44$ & & $0.0916 \pm 0.0071$ & & $\bf{0.474 \pm 0.034}$
\\
WiggleZ & & $0.6$ & & $0.0726 \pm 0.0034$ & & $\bf{0.442 \pm 0.020}$
\\
WiggleZ & & $0.73$ & & $0.0592 \pm 0.0032$ & & $\bf{0.424 \pm 0.021}$\\\\
\hline \hline
\end{tabular}
\end{center}
\end{table}
%\end{widetext}
Then, the contribution of 
BAO to the $\chi^2$ leads to
\be
\chi^2_{BAO}=\bf{X_{BAO}}^{T}{\bf C_{BAO}}^{-1}\bf{X_{BAO}}\ ,
\label{chiBAO}
\ee 
where $\bf{X_{BAO}}$ is the difference vector among the observational data
in Table \ref{table3} and the theoretical points $\{d_z i$ and $A_i (z)\}$,
\be
{\bf{X_{BAO}}}^{\bf{T}} = \left( d_z1 - 0.336,\; d_z2 - 0.1905,\; d_z3 - 0.1097,\; A_4(z) - 0.474,\; A_5(z) - 0.442,\; A_6(z) - 0.424 \right)\ .
\label{Xbao}
\ee 
Hence, by minimizing $\chi^2 = \chi_{CMB}^2 + \chi_{BAO}^2$, the best fit for the above parametrizations is found. In the case of $\Lambda$CDM, the best fit leads to 
$\Omega_{0m}=0.249 \pm 0.009$ and $\Omega_{b}=0.0428 \pm 0.001$ which are then 
used for the calculation of $\chi^2$ for the models (\ref{2.1}) and (\ref{2.2}). 
The results are shown in Table \ref{table4} and Figure \ref{fig:SR.1}, where the best fit of the parameters $\{w_0, z_0\}$ is quite close to the previous one obtained by using Sne Ia data, or at least within the confidence region. The large indetermination of the parameter $z_0$ is also appreciated because of the same point as above.\\
 %And regarding singularities, we have a similar results as in section \ref{SnIa}, both models have a possible singularity within the $1 \sigma$ confidence region.
%\begin{widetext}
\vspace{-2pt}
\begin{table}[h!]  %%%%%%%%% results for BAO+CMB %%%%%%%%%%%
\begin{center}
\begin{minipage}{0.88\textwidth}
\caption{Best fit for the models (\ref{2.1}) and (\ref{2.2}) by using BAO and CMB data, where $\Omega_{0m}=0.249$ and $\Omega_{b}=0.0428$ are assumed. The $\Lambda$CDM model gives $\chi^2_{min}=4.33$ and $ \chi^2_{red}=0.619$. \label{table4}}
\end{minipage}\\
\begin{tabular}{ccccccccc}
\hline
\hline
\bf{Models} & & $\bf{\chi^2_{min}}$ & & $\bf{w_0}$ & & $\bf{z_0}$ & & $\chi^2_{red}$\\
\hline %\vspace{-1pt}\\
$w_1$ & & $4.36$ & & $ -0.09 \pm 0.1$ & & $ -18 \pm 10$ & & $ 0.872$
\\
$w_2$ & & $4.31$ & & $-0.009 \pm 0.06$ & & $ 0.689654 \pm 14$ & & $ 0.862$\\
\hline \hline
\end{tabular}
\end{center}
\end{table}
%\end{widetext}
%\begin{widetext}
\vspace{-2pt}
\begin{table}[Hh!]  %%%%%%%%% results for combined chi^2 %%%%%%%%%%%
\begin{center}
\begin{minipage}{0.88\textwidth}
\caption{Best fit for the models (\ref{2.1}) and (\ref{2.2}) by combining Sne Ia data and standard rulers. The best fit for $\Lambda$CDM  is also shown.  \label{table5}}
\end{minipage}\\
\begin{tabular}{ccccccccccc}
\hline
\hline
\bf{Model} & & $\bf{\chi^2_{min}}$ & & $\bf{\Omega_{0m}}$ & & $\bf{w_0}$ & & $\bf{z_0}$ & & $\chi^2_{red}$\\
\hline %\vspace{-1pt}\\
$w_1$ & & $545.3$ & & $0.250 \pm 0.002$ & & $-0.006 \pm 0.03 $ & & $ 8 \pm 60$ & & $ 0.970$\\
$w_2$ & & $544.5$ & & $0.253 \pm 0.005$ & & $-0.03 \pm 0.02$ & & $ 40 \pm 70$ & & $ 0.969$\\
$\Lambda$CDM & & $548.1$ & & $0.250 \pm 0.005$ & & $-$ & & $-$ & & $ 0.972$\\
\hline \hline
\end{tabular}
\end{center}
\end{table}
%\end{widetext}

%%%%%%%%%%%%%%%%%%%%%%%%%%%%%%%%%%%%%%%%%
%\section{Combining data from SN Ia and Standard Rulers}
%\label{CD}
%%%%%%%%%%%%%%%%%%%%%%%%%%%%%%%%%%%%%%%%%

Finally, let us combine both datasets from previous analysis in order to get a better fit of the free parameters. The resulting grid of the free parameters $\chi(\Omega_m^0, w_0, z_0)$, is analyzed where now $\Omega_m^0$ is kept as a free parameter. The results are shown in Table \ref{table5}, where the best fit of the $\Lambda$CDM model is also included, whereas the resulting contour plots are depicted in Figs. \ref{fig:C.1}-\ref{fig:C.2}. As shown, the best fit for the free parameters are within the confidence regions analyzed previously, whereas the reduced $\chi^2_{red}$ is smaller than the one obtained for the $\Lambda$CDM model. Nevertheless, the indetermination of the parameter $z_0$ remains very large as well as the error on the $w_0$ parameter in both models. In addition, the best fit for all the models gives a similar value for the relative matter density $\Omega_m^0\sim 0.25$, which state the high dependence of the cosmological evolution on matter density independently of the dark energy EoS.

%We see that this minimum is consistent with the previous results, but in both previous computations with Standard Ruler and SNe Ia data, the parameter $z_0$ is quite different comparing to the one obtained with the combined data.\\

%As comparison, the same computation has done for the $\Lambda$CDM model, which give us the following result.
%\vspace{-2pt}
%\begin{table}[h!]  %%%%%%%%% results for combined chi^2 %%%%%%%%%%%
%\begin{center}
%\begin{minipage}{0.88\textwidth}
%\caption{Results of combining the two dataset for the $\Lambda$CDM model. \label{table7}}
%\end{minipage}\\
%\begin{tabular}{ccccccc}
%\hline
%\hline
%\bf{Model} & & $\bf{\chi^2_{min}}$ & & $\bf{\Omega_{0m}}$ & & $\chi^2_{red}$\\
%\hline %\vspace{-1pt}\\
%$\Lambda$CDM & & $548.1$ & & $0.250 \pm 0.005$ & & $ 0.972$\\
%\hline \hline
%\end{tabular}
%\end{center}
%\end{table}

%%%%%%%%%%%%%%%%%%%%%%%%%%%%%%%%%%%%%%%%%%%%%%%%
\section{Discussions}
\label{Conclusions}
%%%%%%%%%%%%%%%%%%%%%%%%%%%%%%%%%%%%%%%%%%%%%%%%
In this paper, we have focused on the analysis of some parametrizations of the dark energy EoS, where we have implemented a new method to reconstruct a scalar field Lagrangian that gives rise to a particular EoS parameter.  Parametrizing the dark energy EoS is commonly used to describe effectively the underlying theoretical model, since it  facilitates the analysis of the dark energy EoS and its confrontation with the observational data. Within this aim we have focused on the reconstruction of a simple theoretical model,  a (non)canonical scalar field, starting from the EoS parameter, where several examples have been studied. Then,  two new parametrizations have been proposed, analyzed in terms of a scalar field and compared with the observational data. The aim of the proposed parametrizations has been to study the possibility of a fast transition and in particular the possibility of crossing the phantom barrier. Note that both models contain $\Lambda$CDM as a special case, when $w_0=0$, in which case the scalar field action reduces to a cosmological constant term, as shown in section \ref{reconstruction}\\
 
Then, by using Supernova Ia and Standard Ruler data, both models have been analyzed and also compared with $\Lambda$CDM.   The results show that both parametrizations leads to a $\chi^2_{min}$ value that is in general slightly smaller than the $\Lambda$CDM model whereas the value of the matter density yields $\Omega_m^0=0.25$ at the best fit. 
%The results show that both parametrizations lead to a $\chi^2_{min}$ value that is in general slightly smaller than the $\Lambda$CDM model whereas the value of the matter density yields $\Omega_m^0=0.25$ at the best fit. 
Besides,  the resulting $\chi^2_{red}$ value within both models is below respect the resulting one for the $\Lambda$CDM model when using both standard candles as standard ruler data. In addition, the second parametrization (\ref{2.2}) leads to better results regarding the value of the $\chi^2_{min\ (red)}$ in comparison with the parametrization (\ref{2.1}).\\

Nevertheless, both parametrizations contain more free parameters than the $\Lambda$CDM model and specifically the parameter $z_0$ presents a large indetermination, specially when dealing with $z_0$ values that can not be constrained with experimental data. Furthermore, by analyzing the different approaches computed along the paper, the contour plots of both models show that $w_0\sim 0$ is very likely, in spite of the best fit is slightly displaced from $\Lambda$CDM.  In addition,  it is remarkable that both models do not lead to future singularities at the best fit, as shown in Table \ref{table5}, and neither in most of the previous results, although the possibility of the occurrence of a 
future singularity is not excluded within the confidence region of the contour plots, as depicted in Figs.~\ref{fig:SN.1}-\ref{fig:C.2}. \\

 Indeed, while analyzing the first model with Sne Ia, the best fit yields $w_0$ very close to $0$ and Fig.~\ref{fig:SN.1} shows that a phantom transition is not excluded but unlikely. The fit with Standard Rulers leads to a similar result as well as when combining both datasets, as shown in Fig.~\ref{fig:C.1}, where specially the $w_0-\Omega_m^0$ contour plot favors $w_0\sim 0$. Nevertheless, the best fit  for $w_1$ shows that the EoS parameter  tends  to an effective cosmological constant in the past ($z \gg 0$) whereas ends up slightly above the phantom barrier at small redshifts. Regarding the second parametrization (\ref{2.2}),  the $w_0$ parameter is 
always negative at the best fit  independently of the data source used, which gives rise to an EoS parameter $w_2$ that  crosses the phantom  barrier at large redshifts  $z\gg z_0$ for the best fit, whereas remains above 
the phantom barrier at small redshifts. In addition, $w_0\sim 0$ is also favored, as shown in the right panels of Figs.~\ref{fig:SN.1}-\ref{fig:SR.1} and Fig.~\ref{fig:C.2}, but a phantom transition is within the confidence region of the contour plots, specifically the best fit leads to a phantom dark energy fluid in the past that tends to a non-phantom regime at small redshifts as pointed out above.\\

On the other hand,  it is remarkable that neither models leads to future singularities at the best fit, as shown in Table \ref{table5}, although the possibility of the occurrence of a 
future singularity is not excluded within the confidence region of the contour plots, as depicted in Figs.~\ref{fig:SN.1}-\ref{fig:C.2}.\\

Hence, the analysis of the present manuscript shows that dealing with effective descriptions of the dark energy EoS can be well connected with the reconstruction of the underlying theory. In addition, both new parametrizations studied  here show that a phantom epoch is compatible with the observational data in spite of that  $\Lambda$CDM model is still very likely, although the best fit deviates a little bit from a cosmological constant. Moreover, the analysis shows that a singularity in the future is not excluded but also unlikely in both parametrizations.
% the experimental points used in this work do not show an inevitable phantom phase, as this behavior is allowed by the parametrizations but avoids a phantom epoch at $z=0$, and seems to be close to a cosmological constant EoS in the far future. Then, within these parametrizations, a singular scenario in the universe evolution seems little probable but a phantom phase may have already occurred. This result may indicate that the proposed models may cross the phantom phase, but will not lead in general to any future singularity. 

\section*{Acknowledgments}
 
We would like to thank Jacobo Asorey, Ruth Lazkoz, Vincenzo Salzano, Irene Sendra and Jon Urrestilla for useful comments and discussions about the manuscript. We also thank the referee of this paper for comments and criticisms that led to a great improvement.  I. L. acknowledges a PIF fellowship from the University of the Basque Country. D. S.-G. acknowledges the support from the University of the Basque Country, Project Consolider CPAN No. CSD2007-00042, the NRF financial support from the University of Cape Town (South Africa) and MINECO (Spain) project FIS2010-15640.

\section*{Appendix}
\label{appendix}

Here we reconstruct explicitly the scalar field Lagrangian by approximating the EoS's  (\ref{2.1}) and (\ref{2.2}) by Pad\'e expansions, specifically the Pad\'e approximation to the exponential function is used, which is given by
\be 
\exp(z) \approx  \dfrac{p_{n}(z)}{p_{n}(-z)}\ ,
\ee
where 
\be 
p_{n}(z) = \sum_{j=0}^n \frac{ \binom {a}{b}}{ j ! \binom {a}{b}} z^j \; .
\ee
Then, the function $ \tanh(z)= \dfrac{\exp(z)-\exp(-z)}{\exp(z)+\exp(-z)} $ can be expressed in terms of a Pad\'e series as follows
\be 
\tanh(z) \approx F_{n}(z)=\frac{p_{n}(z)^2-p_{n}(-z)^2}{p_{n}(z)^2+p_{n}(-z)^2} \; .
\label{pade}
\ee
In order to solve exactly the FLRW equation (\ref{ST7}), which is not possible for the EoS's (\ref{2.1}) and (\ref{2.2}), the above Pad\'e approximation is assumed. The series (\ref{pade}) reproduces the behavior of (\ref{2.1}) and (\ref{2.2}) with a great accuracy, and in particular the fast transition that the EoS experiences. Hence, by assuming the $3^{rd}$ order of the Pad\'e approximation (\ref{pade}),
\be 
\tanh(z-z_0) \approx (z-z_0)\frac{ \left(z-z_0\right)^2+15}{6 \left(z-z_0\right)^2+15} \; ,
\ee
the scalar field Lagrangian can be reconstructed. Higher orders in (\ref{pade}) hold the same problem as the exact EoS, but third order is enough accurate. Hence, the Hubble parameter for the parametrization (\ref{2.1}) by using the Pad\'e approximation is given by
\be
H(z)=A(z)  \e^{B(z)} \sqrt{C(z)}\ , \nonumber
\ee
where
\bea
A(z)&=&H_0\left(2 z^2-4 z z_0+2 z_0^2+5\right)^{-\frac{25 w_0 (z_0+1)}{8 \left(2 z_0^2+4
   z_0+7\right)}}\ , \nn
B(z)&=&\frac{25 w_0 (z_0+1) \log \left[2 (z-z_0)^2+5\right]-4 w_0 \left\{z_0 \left[z_0 (z_0+9)+30\right]+37\right\}
   \log (z+1)}{8 [2 z_0 (z_0+2)+7]}\ , \nn
  C(z)&=&c_1 \left[2 (z-z_0)^2+5\right]^{\frac{25 w_0 (z_0+1)}{4 (2 z_0 (z_0+2)+7)}} \exp\left\{ \frac{1}{4}
   w_0\left[\frac{25 \sqrt{10} \tan ^{-1}\left(\sqrt{\frac{2}{5}} (z-z_0)\right)}{2 z_0 (z_0+2)+7}+2
   z+2\right]\right\}\nn
   &&+\Omega_m (z+1)^{\frac{w_0 \{z_0 [z_0 (z_0+9)+30]+37\}}{2 z_0
   (z_0+2)+7}+3}\ .
   \label{HP1}
 \eea
Note that the approximation (\ref{pade}) at third order fits the EoS (\ref{2.1}) very accurately around the $z_{0}$, while far from this point, the deviation grows linearly by $\sim6 \% $ every $100z$. Nevertheless, the main feature of the above EoS (a fast transition at $z=z_0$) is greatly achieved by (\ref{pade}), while far away from $z_0$, the EoS (\ref{2.1}) becomes constant (\ref{s.2}). Then, the corresponding scalar field Lagrangian is reconstructed, where the kinetic term yields
\be
\gamma(\phi)=\frac{c_1 w_0}{\kappa^2  \phi ^2 }\frac{ a(\phi)\e^{b(\phi)}}{c(\phi) \e^{b(\phi)}+d(\phi)}\ , \nonumber
\ee 
where
\bea
a(\phi) &=& \left\{-\left[z_0 (z_0 (z_0+9)+30)+37\right] \phi ^3+3 \left[z_0 (z_0+6)+10\right] \phi ^2-3
   (z_0+3) \phi +1\right\}\nn
   &\times& \left\{\frac{\left[2 z_0 (z_0+2)+7\right] \phi ^2-4 (z_0+1) \phi +2}{\phi
   ^2}\right\}^{\frac{25 w_0 (z_0+1)}{4 (2 z_0 (z_0+2)+7)}-1}\ ,\nn
b(\phi)&=&\frac{1}{4} w_0
   \left\{\frac{2}{\phi }-\frac{25 \sqrt{10} \tan ^{-1}\left[\frac{\sqrt{\frac{2}{5}} (z_0 \phi +\phi -1)}{\phi }\right]}{2z_0 (z_0+2)+7}\right\}\ ,\nn
c(\phi)&=&c_1 \phi ^3 \left\{\frac{[2 z_0 (z_0+2)+7] \phi
   ^2-4 (z_0+1) \phi +2}{\phi ^2}\right\}^{\frac{25 w_0 (z_0+1)}{4 (2 z_0 (z_0+2)+7)}}\ ,\nn
d(\phi)&=&\Omega_m \left(\frac{1}{\phi}\right)^{\frac{w_0 \{z_0 [z_0 (z_0+9)+30]+37\}}{2 z_0
   (z_0+2)+7}}\ .
\label{HP2}
   \eea
While the potential scalar leads to
\be
V(\phi)=-\frac{c_1 H_0^2}{2 \kappa^2}f(\phi) \e^{j(\phi)}\ , \nonumber
 \ee  
 where
 \bea
 f(\phi)&=& \left\{\phi ^3 \left[-(w_0 (z_0 (z_0 (z_0+9)+30)+37)+6 (2 z_0 (z_0+2)+7))\right]+3
   \phi ^2 (w_0 (z_0 (z_0+6)+10)+8 (z_0+1))\right.\nn
&&\left. -3 \phi  (w_0 (z_0+3)+4)+w_0\right\} \times\left[\frac{2 (z_0 \phi +\phi -1)^2}{\phi ^2}+5\right]^{\frac{25 w_0 (z_0+1)}{4 (2 z_0
   (z_0+2)+7)}-1}\left(\frac{1}{\phi }\right)^{3-\frac{w_0 (z_0 (z_0 (z_0+9)+30)+37)}{2 z_0(z_0+2)+7}}\ , \nn
   j(\phi)&=&\frac{1}{4} w_0 \left\{\frac{25 \sqrt{10} \tan ^{-1}\left[\sqrt{\frac{2}{5}}
   \left(-z_0+\frac{1}{\phi }-1\right)\right]}{2 z_0 (z_0+2)+7}+\frac{2}{\phi }\right\}\ .
   \label{HP3}
\eea
In the same way, let us now consider the second parametrization (\ref{2.2}). As above, by considering the Pad\'e approximation at third order, the FLRW equation is solved exactly and the resulting Hubble parameter yields
\be
H(z)=A(z)\exp^{B(z)} \sqrt{C(z)}\ , \nonumber
\ee
where,
\bea
A(z)&=&H_0\left(2 z^2-4 z z_0+2 z_0^2+5\right)^{-\frac{25 w_0 (z_0+1)}{8 \left(2 z_0^2+4
   z_0+7\right)}}\ , \nn 
B(z)&=&\frac{w_0 (z_0+1) \left\{25 \log \left[2 (z-z_0)^2+5\right]-4 \left[z_0
   (z_0+2)+16\right] \log (z+1)\right\}}{8 (2 z_0 (z_0+2)+7)}\ , \nn
C(z)&=&c_1 \left[2
   (z-z_0)^2+5\right]^{\frac{25 w_0 (z_0+1)}{4 (2 z_0 (z_0+2)+7)}} \exp \left\{\frac{1}{4}
   w_0 \left[\frac{25 \sqrt{10} \tan ^{-1}\left(\sqrt{\frac{2}{5}} (z-z_0)\right)}{2 z_0 (z_0+2)+7}+2
   z+2\right]\right\}\nn
   &&+\Omega_m (z+1)^{\frac{w_0 (z_0+1) [z_0 (z_0+2)+16]}{2 z_0
   (z_0+2)+7}+3}\ .
   \label{HP4}
\eea
Then, the scalar Lagrangian for the parametrization $w_2$ can be reconstructed, where the kinetic term leads to
\be
\gamma(\phi)=\frac{c_1 w_0}{\kappa^2  \phi ^2}\frac{a(\phi)\e^{b(\phi)}}{c_1 \phi ^3 c(\phi)\e^{b(\phi)}+d(\phi)}\ , \nonumber
%   \label{HP5}
 \ee
 where
 \bea
 a(\phi)&=&-(z_0 \phi +\phi -1) \left\{[z_0 (z_0+2)+16] \phi ^2-2 (z_0+1) \phi +1\right\}
   \left\{\frac{[2 z_0 (z_0+2)+7] \phi ^2-4 (z_0+1) \phi +2}{\phi ^2}\right\}^{\frac{25 w_0
   (z_0+1)}{4 [2 z_0 (z_0+2)+7]}-1}\ , \nn
b(\phi)&=&\frac{1}{4} w_0 \left\{\frac{2}{\phi }-\frac{25 \sqrt{10}
   \tan ^{-1}\left[\frac{\sqrt{\frac{2}{5}} (z_0 \phi +\phi -1)}{\phi }\right]}{2 z_0
   (z_0+2)+7}\right\}\ , \nn
 c(\phi)&=& \left\{\frac{[2 z_0 (z_0+2)+7] \phi ^2-4
   (z_0+1) \phi +2}{\phi ^2}\right\}^{\frac{25 w_0 (z_0+1)}{4 (2 z_0 (z_0+2)+7)}}\ , \nn
  d(\phi)&=&\Omega_m \left(\frac{1}{\phi
   }\right)^{\frac{w_0 (z_0+1) [z_0 (z_0+2)+16]}{2 z_0 (z_0+2)+7}}\ .   
\label{HP5}
   \eea
And the scalar potential yields
\be
V(\phi)=\frac{c_1H_0^2}{2 \kappa^2 }f(\phi) \e^{j(\phi)}
\ee
where
\bea
f(\phi)&=&\left\{\left(-z_0+\frac{1}{\phi }-1\right) \left[w_0 \left(\left(z_0-\frac{1}{\phi
   }+1\right)^2+15\right)+12 \left(z_0-\frac{1}{\phi }+1\right)\right]-30\right\} \left[\frac{2 (z_0
   \phi +\phi -1)^2}{\phi ^2}+5\right]^{\frac{25 w_0 (z_0+1)}{4 (2 z_0 (z_0+2)+7)}-1}  \nn
   &&\times \left(\frac{1}{\phi
   }\right)^{-\frac{w_0 (z_0+1) (z_0 (z_0+2)+16)}{2 z_0 (z_0+2)+7}}\ , \nn
j(\phi)&=&\frac{1}{4} w_0 \left\{\frac{25 \sqrt{10} \tan ^{-1}\left[\sqrt{\frac{2}{5}} \left(-z_0+\frac{1}{\phi
   }-1\right)\right]}{2 z_0 (z_0+2)+7}+\frac{2}{\phi }\right\}\ .
\label{HP6}
\eea
Hence, the parametrizations analyzed along this work can be constructed analytically in terms of the scalar field Lagrangian (\ref{ST1}) by using Pad\'e series.

\end{document}